\documentclass[a4paper,twocolumn]{aa} 
\usepackage{graphicx}
\usepackage{txfonts}
\usepackage{natbib}
\bibpunct{(}{)}{;}{a}{}{,} 
\begin{document}
\newcommand{\masq}{magn \arcsec$^{-2}$}
\newcommand{\Jykms}{Jy\,km\,s$^{-1}$}
\newcommand{\mjyb}{mJy beam$^{-1}$}
\newcommand{\HO}{$H_0$}
\newcommand{\kms}{km\,s$^{-1}$}
\newcommand{\acm}{cm$^{-2}$}
\newcommand{\teff}{$T_{eff}$} 
\newcommand{\kmsmpc}{km\,s$^{-1}$\,Mpc$^{-1}$}
\newcommand{\mjb}{mJy beam$^{-1}$}
\newcommand{\jb}{Jy beam$^{-1}$}
\newcommand{\mjbc}{mJy beam$^{-1}$ channel$^{-1}$}
\newcommand{\jykms}{Jy km s$^{-1}$}
\newcommand{\msun}{$\cal M_\odot$}
\newcommand{\msunyr}{$\cal M_\odot$\,yr$^{-1}$}
\newcommand{\lsun}{{$L_{\odot,B$}}}
\newcommand{\lb}{{L$_{\rm B}$}}
\newcommand{\mlsun}{{\rm M}$_\odot/{\rm L}_{\rm B_{\odot}$}}
\newcommand{\kmsMpc}{km s$^{-1}$ Mpc$^{-1}$}
\newcommand{\hi}{H{\small I}}
\newcommand{\MHI}{{$\cal M}_{\rm HI}$}
\newcommand{\m}{\hbox{$^{\rm m}$}}
\newcommand{\s}{\hbox{$^{\rm s}$}}
\newcommand{\h}{\hbox{$^{\rm h}$}}
\newcommand{\dn}{D$_{\rm n}$4000}
\newcommand{\halpha}{H$\alpha$}
\newcommand{\ewhalpha}{EW(H$\alpha$)}
\newcommand{\hbeta}{H$\beta$}
\newcommand{\ewhbeta}{EW(H$\beta$)}
\newcommand{\hgamma}{H$\gamma$}
\newcommand{\hdelta}{H$\delta$}
\newcommand{\hdeltaA}{H$\delta_{A}$}
\newcommand{\ewhdelta}{EW(H$\delta$)}
\newcommand{\hds}{H$\delta$~Strong}
\newcommand{\sn}{\small S/N}
\newcommand{\oii}{[O{\small II}]}
\newcommand{\ewoii}{EW[O{\small II}]}
\newcommand{\oiii}{[O{\small III}]}
\newcommand{\ewoiii}{EW[O{\small III}]}
\newcommand{\neiii}{[Ne{\small III}]} 
\newcommand{\ie}{{i.e.\,}}
\newcommand{\eg}{{e.g.\,}}
\newcommand {\LCDM} {\ifmmode \Lambda{\rm CDM} \else $\Lambda{\rm CDM}$ \fi}
\newcommand{\MASSIV}{\small MASSIV}
\newcommand{\massiv}{\small MASSIV}
\newcommand{\images}{\small IMAGES}
\newcommand{\IMAGES}{\small IMAGES}

\title{{MASSIV: Mass Assembly Survey with SINFONI in VVDS}
  \thanks{This work is based mainly on observations collected at the
    European Southern Observatory (ESO) Very Large Telescope, Paranal,
    Chile, as part of the Programs 179.A-0823, 177.A-0837, 78.A-0177,
    75.A-0318, and 70.A-9007. This work also benefits from data
    products produced at TERAPIX and the Canadian Astronomy Data
    Centre as part of the Canada-France-Hawaii Telescope Legacy
    Survey, a collaborative project of NRC and CNRS.}}
\subtitle{IV. Fundamental relations of star-forming galaxies at $1 < z
  < 1.6$\thanks{All the data published in this paper are publicly
    available following this link:
    http://cosmosdb.lambrate.inaf.it/VVDS-SINFONI}}
  \titlerunning{Fundamental relations from Near-infrared resolved
    kinematics}

\author{
D. Vergani \inst{1,2}
\and B. Epinat\inst{3,4,5}
\and T. Contini\inst{3,4}
\and L. Tasca\inst{5}
\and L. Tresse\inst{5}
\and P. Amram\inst{5}
\and B. Garilli\inst{6}
\and M. Kissler-Patig\inst{7}
\and O. Le F\`{e}vre\inst{5}
\and J. Moultaka\inst{3,4}
\and L. Paioro\inst{6}
\and J. Queyrel\inst{3,4}
\and C. L\'{o}pez-Sanjuan\inst{5}
}
\offprints{\mbox{D.~Vergani}, \email{\it vergani@iasfbo.inaf.it}}
\institute{
{INAF-IASFBO, Via P. Gobetti 101, I-40129 Bologna, Italy, \email{vergani@iasfbo.inaf.it}} \and
{INAF-Osservatorio Astronomico di Bologna, Via C. Ranzani 1, I-40127 Bologna, Italy} \and
{Institut de Recherche en Astrophysique et Plan\'etologie (IRAP), CNRS, 14 Avenue \'Edouard Belin, F-31400 Toulouse, France}\and
{IRAP, Universit\'e de Toulouse, UPS-OMP, F-31400 Toulouse, France}\and
{Laboratoire d'Astrophysique de Marseille, Universit\'e d'Aix-Marseille, CNRS, 38 rue Fr\'ed\'eric Joliot-Curie, F-13388 Marseille Cedex 13, France} \and
{INAF-IASFMI, Via E. Bassini 15, I-20133 Milano, Italy} \and
{ESO, Karl-Schwarzschild-Str. 2, D-85748 Garching b. M\"unchen, Germany}
}
\date{Received 14 November 2011 / accepted 14 February 2012}

\abstract {}{How mass assembly occurs in galaxies and which
  process(es) contribute to this activity are among the most highly
  debated questions in galaxy formation and evolution theories. This
  has motivated our survey {\small MASSIV} (Mass Assembly Survey with
  SINFONI in VVDS) of $0.9 < z < 1.9$ star-forming galaxies selected
  from the purely flux-limited VVDS redshift survey.}  {We evaluate
  the characteristic size and stellar mass of 46 {\small MASSIV}
  galaxies at $1 < z < 1.6$ and use the internal dynamics obtained
  with the SINFONI integral field spectrograph mounted at the Very
  Large Telescope.} 
{For the first time, we derive the relations between galaxy size,
  mass, and internal velocity, and the baryonic Tully-Fisher relation,
  from a statistically representative sample of star-forming galaxies
  at $1 < z < 1.6$. We find a dynamical mass that agrees with those of
  rotating galaxies containing a gas fraction of $\sim 20$\%, that is
  perfectly consistent with the content derived using the
  Kennicutt-Schmidt formulation and corresponds to the expected
  evolution. Non-rotating galaxies have more compact sizes for their
  stellar component, and are less massive than rotators, but do not
  have statistically different sizes for their gas-component. Discs of
  ionized gas have irregular, clumpy distributions, but the simplistic
  assumption of exponential profiles is verified. We measure a
  marginal evolution in the size - stellar mass and size - velocity
  relations in which discs become evenly smaller with cosmic time at
  fixed stellar mass or velocity, and are less massive at a given
  velocity than in the local Universe. This result is inconsistent
  with previous reports of an abnormal evolution in the galactic
  spin. The scatter in the Tully-Fisher relation is smaller when we
  introduce the $S_{05}$ index, which we interpret as evidence of an
  increase in the contribution to galactic kinematics of turbulent
  motions with cosmic time. We report a persistently large scatter for
  rotators in our relations, that we suggest is intrinsic, and
  possibly caused by complex physical mechanism(s) at work in our
  stellar mass/luminosity regime and redshift range.} {Our results
  consistently point towards a mild, net evolution of these relations,
  comparable to those predicted by cosmological simulations of disc
  formation. In a conflictual picture where earlier studies reported
  discrepant results, we place on firmer ground the measurement of a
  lack of any large evolution of the fundamental relations of
  star-forming galaxies in at least the past 8~Gyr and evidence that
  dark halo is strongly coupled with galactic spectrophotometric
  properties.}  \keywords{Galaxies: high-redshift - galaxies:
  evolution - galaxies: kinematics and dynamics - galaxies:
  fundamental parameters } \authorrunning{D. Vergani et al.}
\titlerunning{Fundamental relations at $z\sim 1.2$} \maketitle

\section{Introduction}
\label{sec:intro}

In the current paradigm of structure formation, galaxies are formed in
dark matter halos that grow with cosmic time via the hierarchical
assembly of smaller units. While halo$-$halo merging is the main
physical process driving the assembly of dark$-$matter halos and
successfully describes the galaxy clustering properties, it is unclear
how it affects the build$-$up of galaxies, and how other physical
processes might play a role.

Large and deep spectroscopic surveys of local and higher$-z$ galaxies
have been performed in the past decade to improve our understanding of
galaxy formation and evolution. Among the many trends that have been
found between various galaxy properties (e.g., luminosity, colour,
stellar mass, surface brightness, and star$-$formation rate), a
general consensus has been reached on a luminosity/mass$-$dependent
evolution of the distribution functions
\citep[\eg][]{zuc06,poz07,bun06,fon04,fon06} supporting the downsizing
scenario \citep[\eg][]{cow96}.

There is convincing evidence that $1 < z < 2.5$ is a critical cosmic
period, when galaxy properties changed rapidly. During this period, we
understand that there was an intense mass assembly activity and that
the morphological distinction between galaxies became
established. This epoch corresponds to the maximal star formation
activity in the Universe around $z\sim 2$
\citep[e.g.][]{tre07,cucciati2011}, and the rapid build$-$up of the
stellar mass observed in early$-$type galaxies at $z \sim 1-1.5$
\citep{arn07,ver08}. It is also an epoch when the connection between
AGN activity and star formation might have played an important role in
shaping galaxies. In this framework, knowledge of the dynamical state
of galaxies is indispensable to assessing the mechanisms governing the
growth of luminous structures in the Universe.

In the past few years, the availability of integral field
spectrographs working under excellent seeing conditions, or assisted
by adaptive optics, has opened a new era in the investigation of
galactic resolved dynamics at high redshifts, with integral field
spectrographs providing considerably more information than traditional
long-slit techniques. Acquiring the full two-dimensional maps of both
the velocity fields and velocity dispersions is far more helpful to
many dynamical studies than the classical long-slit technique because
it avoids any misinterpretation of the measured velocities due to
either the misalignment between the photometric and kinematic axes, or
large velocity dispersions. This is especially true at increasingly
large look-back times when galaxies display a larger incidence of
irregular shapes and chaotic motions. In addition, new opportunities
have recently been arisen in the near-infrared regime in which new
instruments allow one to sample the rest-frame optical emission that
contains all major diagnostic lines, e.g., SINFONI on the Very Large
Telescope (VLT), \citet{eisenhauer03} and OSIRIS on Keck,
\citet{lar06}. Spatially-resolved observations of UV/optically
selected galaxies at $1.5<z<3$ \citep{genzel06, genzel08, nfs06,
  nfs09, bouche07, law07,law09, wright2008, stark08, epinat2009}
together with cosmological simulations \citep[e.g.,][]{genel08,
  dekel09} indicate that many of the physical properties of these
sources cannot be primarily attributed to major merger events
experienced in the history of galaxies. Internal mechanisms instead
contribute to mass assembly \citep[][ and references
 therein]{genzel08}. In determining the observed dynamics, these
processes might correspond to the infall of accreting matter, and/or
large collisional clumps \citep[e.g.,][]{law07, law09,genzel2011}
and/or processes related to star formation \citep{lehnert2009}.

In summary, the emerging scenario places a far smaller importance on
major merging in shaping up galaxies at early epochs, apparently at
odds with the hierarchical assembly of mass in the cold dark matter
paradigm - although merging systems have been identified in about one
third of the galaxies studied. A general consensus has yet to be
reached \citep[e.g.,][]{robertson08}.

To date a detailed view of dynamics of distant galaxies has been
restricted to a limited number of objects primarily because of the
large investment of telescope time required to build up a fair
sample. The most efficient approach has been to adopt certain
selection techniques, such as various combinations of colours to focus
on a specific redshift range and/or target extended discs to ensure
sufficient spatial sampling \citep{law07, law09,nfs06, nfs09}. While
pre-selecting galaxies remains the most effective way to ensure the
production of a large, high-$z$ galaxy catalogue, when exploring
statistical issues it is highly desirable to have a fully
representative catalogue of galaxies with well-known selection
functions in multiple fields of a homogeneous survey
\citep[e.g.,][]{puech10}. The importance of a sufficient spatial
resolution has also been demonstrated by surveys such as IMAGES
(Intermediate MAss Galaxy Evolution Sequence), which selects $z\sim
0.6$ galaxies observed with limited spatial resolution (0\,\farcs52
pix$^{-1}$). The resulting picture emerging from various data-sets
provides discrepant results about the evolution of the fundamental
relations between galaxy size, mass, and velocity
\citep[e.g.,][]{flores06, puech08, puech10, bouche07, nfs09, cresci09,
  law07, law09}. As these quantities represent important indicators
of the structure and evolution of disc galaxies, the interpretation of
their changes, or constancy with time, helps us to determine the
standard picture of galaxy disc formation. For instance, at odds with
earlier findings and theoretical expectations, the absence of any
evolution in the size-velocity relation of disc galaxies, as reported by
Bouch{\'e} et al. (2007), has fundamental implications for the growth
of the specific angular momentum of dark matter halos.

To overcome the aforementioned difficulties and shed light on
relations linking various galactic properties to dynamics, we are
carrying out the {\small MASSIV} programme (Mass Assembly Survey with
SINFONI in VVDS) to obtain near-infrared integral field spectroscopy
of a hundred star-forming galaxies selected by applying a pure
flux-limit criterion to three VVDS fields. The main goal of {\small
 MASSIV} is to obtain a detailed description of galaxy dynamics to
probe their formation and evolution for a representative sample of
galaxies in the crucial $1 < z < 2$ epoch. In the first set of
papers, we present the survey in \citet{contini11}, the kinematic
classification of galaxies in \citet{epinat11}, and the resolved
metallicity data of galaxies in \citet{queyrel11}. In the present
paper, we present the fundamental scaling relations that exist among
the mass, size, and velocity of the studied galaxies.

In Sect.\,2, we summarize our sample selection, observations, and data
reduction. In Sect.\,3, we detail the procedures to determine the
physical properties of the {\small MASSIV} galaxies (morphology, mass,
star formation rate, and kinematics inferred from the photometric and
integral field spectroscopic data). In Sect.\,4, we discuss the gas
contents and sizes of our galaxies, and in Sect.\,5, we analyse our
resulting scaling relations in the context of mass assembly.

Throughout this work, we assume a standard cosmological model with
$\Omega_M = 0.3$, $\Omega_\Lambda = 0.7$, and $H_0 = 70 \, \mathrm{km}
\, \mathrm{s}^{-1} \, \mathrm{Mpc}^{-1}$ (which gives a median scale
of 8.32~kpc~arcsec$^{-1}$ at $z\sim 1.24$). Magnitudes are given in the
AB system.

\begin{table}[bh!]
\caption{Sample selection.}
\tabcolsep0.55mm 
\begin{tabular}{cc rrr r}\hline \hline \\
  VVDS       & $\rm I_{AB,lim}$ & Number~ & Criteria~~               & {\small MASSIV}     & This~~~~~   \\
  Survey fields  & 		 & of targets & {\small Full-filed} & sample~~  & work~~~~  \\
  (1)                  & (2)         & (3)~~~~  & (4)~~~~          & (5)~~~~    & (6)~~~~~~ \\ \\ \hline \\
  Wide                 & 17.5-22.50       & 24,507   & 224 (183)   &  21~ (4)   & 21 (4)~~~~ \\
  Deep                 & 17.5-24.00       & 12,668   & 2,600 (300) &  29~ (1)   & 27 (1)~~~~ \\
  Ultra-Deep           & 23.0-24.75       &  1,200   & 86~~ (55)   &  34~ (6)   &  2 (2)~~~~ \\
  &                    &             &          &                  &                   \\
  \hline \hline
\label{tab:fields}
\end{tabular}
\\ { (1) The fields of the VVDS used in {\small MASSIV}. (2) ${\rm
   I_{AB}}$ limiting magnitude selection. (3) Total number of
  spectroscopic targets. (4) Number of targets fulfiling the criteria
  described in Sect.\,\ref{sec:sample}. (5) Total number of galaxies
  in the {\small MASSIV} sample at the completion of the survey. (6)
  Number of galaxies used in this analysis. In columns (4) to (6), we indicate in parentheses the numbers of targets observed with adaptive optics in our observing campaign.}
\end{table}

\section{Observations}
\subsection{Sample selection}
\label{sec:sample}

We report SINFONI \citep{eisenhauer03, bonnet04} observations of our
new resolved kinematic {\small MASSIV} survey. The observed galaxies
were selected from the VVDS. The VVDS is a redshift survey that has
 presented results on the evolution of galaxies, large-scale
structures, and AGNs based on the spectroscopic identification of about
50,000 sources down to a limiting magnitude of $I_{AB} = 22$ in the
wide fields \citep{gar08}, to $I_{AB} = 24$ in the deep field
\citep{lef05}, and to $I_{AB} = 24.75$ in the ultra-deep field (see
Tab.\,\ref{tab:fields}). The VVDS has been combined with a multi-wavelength
dataset from radio to X-rays.

The {\small MASSIV} sample analysed in this work was selected based on
several criteria (see details in Contini et al. 2012), which can be
summarized as follows:

1) Galaxies with a redshift assigned to their rest-frame UV VVDS
spectra at a confidence level higher than 80\% in the redshift range
$z=0.9-1.8$ were pre-selected. The confidence level of the redshift
determination had been estimated in the VVDS based on repeated
observations. Galaxies were grouped into four classes and we used only
those that corresponded to the maximum success probability of having an
accurate redshift.

2) Their spectral features (\halpha$\lambda$6563\AA\ or
\oiii$\lambda$5007\AA) had to fall in a region not contaminated by OH
sky-lines. Strong sky-lines were outdistanced by a minimum of
9\AA\ from the above-mentioned spectral lines computed using our VIMOS
spectroscopic redshifts. Thus, we could observe an OH-uncontaminated
emission line with a typical rotation of 200~\kms. \smallskip

3) Star-forming galaxies at redshifts $z<1.46$ were selected by taking
a certain threshold in the equivalent width (EW) of the
\oii$\lambda$3727 doublet in VIMOS spectra with a signal-to-noise
ratio (S/N) above 6. Hence, we selected $z<1.46$ galaxies with
\ewoii\ larger than 25\AA\ in S/N$\le$10 spectra and 40\AA\ in $6 \le$
S/N $\le 10$ spectra. Star-forming galaxies at higher redshifts were
selected based on the rest-frame UV spectra. These constraints
guarantee the detection of \halpha$\lambda$6563\AA\ (or
\oiii$\lambda$5007\AA) in SINFONI spectra within the scheduled
observing time. \smallskip

Out of the $\sim 30,000$ spectra observed in the various VVDS fields,
10\% of them satisfy these criteria (see Tab.\,\ref{tab:fields} for
the details). From this subsample, 84 galaxies have been observed in
the framework of {\small MASSIV}. The galaxy VVDS220148046, which had
been assumed to be at redshift 1.3710 based on its VIMOS spectrum,
turns out to have a redshift of 2.2442 identified with the \hbeta\ and
\oiii\ lines in SINFONI observations. This galaxy is not included in
the current analysis. Finally, our 83 galaxies are representative of a
star-forming galaxy population at $0.9 < z < 1.8$ that is uniformly
distributed in the stellar mass interval log${\cal M/M_\odot}
=9.47-11.77$. In this sample, ten galaxies have been observed with the
adaptive optics (AO) system of SINFONI assisted with the Laser Guide
Star (LGS) facility to achieve a higher angular resolution.

In this analysis, we use more than half of the entire {\small MASSIV}
sample, i.e., 50/83 galaxies, or about $\sim 60$\%. Thanks to the high
precision of the VVDS redshifts, almost all observed galaxies have
been detected (45/50) using either the
\halpha$\lambda$6563\AA\ (44/45) or \oiii$\lambda$5007\AA\ (1/45)
spectral lines. In particular, in seeing-limited mode all but one (43
out of 44) of the galaxies targeted in \halpha$\lambda$6563\AA\ have
been detected. Seven out of 50 galaxies have been observed with
adaptive optics (among which two have no detection, and one was
targeted in \halpha\ while another was targeted in \oiii). The 50
galaxies that constitute the original {\small MASSIV} sample (see
Tab.\,1) have a median redshift of $z=1.2423$ (where $z_{25\%}=1.0375$
and $z_{75\%}=1.3547$ are the two extreme quartiles of the redshift
distribution). Excluding our serendipitous detection at $z=2.2442$ and
the four undetected galaxies, the median redshift of our 45 galaxies
analysed in this work remains similar to that of the original sample
($z= 1.2246$ with $z_{25\%}=1.0351$ and $z_{75\%}=1.3338$).

\subsection{SINFONI data acquisition and reduction}
\label{sec:reduction}

Our observations of the {\small MASSIV} sample were collected
with the near-infrared integral field spectrograph SINFONI of the VLT
under the large programme 179.A-0823 (P.I. T. Contini) between April
2007 and January 2011 and include the observations of the pilot runs
75.A-0318 and 78.A-0177. Our subsample of 50 galaxies refers to
observations conducted until 2009 and fully reduced and analyzed
before 2010.

In the accompanying paper by \citet{contini11}, we detail the strategy
and global physical properties of {\small MASSIV} galaxies, along with
a more exhaustive explanation of selection functions. A full description of
the data reduction procedures are given in \citet{epinat11}. Here we
provide a short summary.

We used the J- and H-bands to sample the spectral interval $1.08 -
1.41$~$\mu m$ and $1.43 - 1.86$~$\mu m$ with a spectral resolution of
$R$ $\sim 2000$ and $\sim 3000$, respectively. The majority of our
observations were in seeing-limited mode with a 8\arcsec $\times$
8\arcsec\ field-of-view and 0\farcs125 $\times$ 0\farcs25 pixel scale.
For a subsample of seven galaxies, we obtained AO observations with a
3\farcs2 $\times$ 3\farcs2\ field-of-view and 0\farcs100 $\times$
0\farcs050 pixel scale. To maximize the telescope time, we placed the
majority of our targets at different positions in the field-of-view of
the instrument. This strategy avoids the need for sky frame
acquisition (for detail see in Contini et al. 2012). In addition, we
applied a sub-dithering to place the target at different positions on
the chip. Conditions were photometric during the observations with a
median seeing of 0\,\farcs70 measured using observations of the stars
acquired to measure the point spread function (PSF) during the
night. To accurately point our galaxies, we acquired them by
performing a blind offset from a bright nearby star. We also observed
standard stars over the same nights to enable us to flux-calibrate our
spectra. Individual exposures ranged between 300s and 900s with a
total on-source integration time that ranged between 1h and 2h.

The core of the data reduction was performed using the ESO-SINFONI
pipeline, version 2.0.0 \citep{Modigliani2007} complemented with
additional routines to homogenize the data processing among the
different reducers. We subtracted the sky background, skylines, and
dark current from the raw science data using the contiguous sky frames
obtained with the offset-to-sky sequence. The data were flat-fielded
using observations of an internal lamp that had been wavelength
calibrated with spectra of reference arc-lamps. These processed data
frames were then reconstructed into data cubes. Individual cubes of a
given observing block were aligned in the spatial direction by relying
on the telescope offsets from a nearby bright star, and then
combined. The rejection of cosmic rays was applied to the combined
cubes. The observations of a telluric standard star of a typical B
spectral type were acquired immediately after each science frames and
were reduced in a similar way. We extracted the mono-dimensional
spectra of the stars by summing the flux of an aperture of a diameter
of five resolution elements to derive the flux calibration after
removing the telluric absorption lines. We corrected for the
atmospheric transmission dividing the science cubes by the integrated
spectrum of the telluric standard. We created sky cubes to quantify
the effective resolution for each science data cube. The sky cubes
were reduced in a similar way to the science frames, but no correction
for sky subtraction was applied. We fitted with Gaussian profiles the
unblended night sky lines in the extracted, mono-dimensional
spectra. The effective FWHM spectral resolution has a typical value of
130~\kms.

\begin{table*}[ht!]
\begin{center} 
\tabcolsep1.4mm 
\caption{\sc {Properties of the {\small MASSIV} sample}}
\begin{tabular}{  c | l l rrr ccc l c r }
  \hline \hline \\
  ID        &  ~~~~$z$  &$Line$           & $M_{*}$~~~~         & $M_{gas}$ & $M_{dyn}$~~~~      &$R_{last}$ & $R_{1/2,gas}$   & $R_{1/2,star}$ & ~~$V_{max}$ & ${V_{max}}/{\sigma}$ & {\small Class}~~~ \\
  (1)       & ~~~(2)   &  (3)            &  ~~~~(4)           & (5)       & (6)~~~~          & (7)~~    &  (8)~~~~~~~   & (9)          & ~~~(10)     & (11)~~~             & (12)~~~ \\  [0.3mm]                                                          
  \hline 
  020106882  &1.3991            &H$\alpha$            &$  9.99^{+0.22}_{-0.22}$  & $9.84$     & $10.41^{+0.18}_{-0.13}$  &   5.1 &   3.79$\pm$0.08     &3.51$\pm$ 0.14 &                   133$\pm$25~ 	  &3.2      & ROT\,    	  \\
  020116027  &1.5302            &H$\alpha$            &$ 10.09^{+0.23}_{-0.23}$  & $9.86$     & $ 9.94^{+0.23}_{-0.29}$  &   6.5 &   3.56$\pm$0.08     &4.27$\pm$ 0.16 &	 	     27$\pm$10~ 	  &0.6      & NON-R\,     \\
  020126402  &1.2332$^{\rm \,a,b}$ &[O${\rm III}$]        &$ 10.09^{+0.26}_{-0.25}$  & $\dots$    & $\dots$             &$\dots$&   $\dots$ 	      &2.12$\pm$ 0.31 &		     $\dots$~             &$\dots$  &$\dots$~     \\
  020147106  &1.5195           &H$\alpha$             &$ 10.10^{+0.13}_{-0.13}$  & $9.89$     & $11.73^{+0.36}_{-0.36}$  &   7.8 &   1.52$\pm$0.08     &1.17$\pm$ 0.41 &		     26$\pm$51~  	  &0.3      & ROT$^{\rm \,e}$\\
  020149061  &1.2905           &H$\alpha$             &$ 10.18^{+0.23}_{-0.23}$  & $9.89$     & $11.29^{+0.80}_{-0.80}$  &   4.8 &   3.68$\pm$0.08     &1.09$\pm$ 0.69 &		     107$\pm$210$^{\rm \,d}$ &1.5      & ROT\,	  \\
  020164388  &1.3547           &H$\alpha$             &$ 10.13^{+0.31}_{-0.31}$  & $9.83$     & $10.77^{+0.26}_{-0.33}$  &   8.2 &   2.78$\pm$2.27     &2.72$\pm$ 0.05 &		     79$\pm$19~  	  &1.5      & NON-R\,	  \\
  020167131  &1.2246           &[O${\rm III}$]$^{\rm c}$&$ 10.08^{+0.19}_{-0.19}$  & $9.81$     &  $9.83^{+0.17}_{-0.10}$  &   1.8 &   3.96$\pm$0.33     &2.60$\pm$ 0.09 &		     127$\pm$29~ 	  &5.0      & NON-R\,     \\
  020182331  &1.2290           &H$\alpha$             &$ 10.72^{+0.11}_{-0.11}$  & $9.92$     & $10.51^{+0.14}_{-0.13}$  &   5.5 &   3.49$\pm$2.08     &4.19$\pm$ 0.17 &		     127$\pm$25$^{\rm \,d}$  &1.9      & NON-R\,     \\
  020193070  &1.0279           &H$\alpha$             &$ 10.15^{+0.20}_{-0.20}$  & $9.57$     & $10.14^{+0.11}_{-0.07}$  &   3.9 &   4.11$\pm$1.37     &3.49$\pm$ 0.09 &		     116$\pm$23$^{\rm \,d}$  &3.6      & NON-R\,     \\
  020208482  &1.0375           &H$\alpha$$^{\rm c}$     &$ 10.17^{+0.16}_{-0.16}$  & $8.94$     &  $9.92^{+0.11}_{-0.08}$  &   1.4 &   5.50$\pm$0.08     &3.67$\pm$ 0.20 &		     158$\pm$31~ 	  &22.9     & ROT\,	  \\
  020214655  &1.0395           &H$\alpha$             &$ 10.02^{+0.16}_{-0.16}$  & $9.90$     & $10.91^{+0.31}_{-0.43}$  &   5.7 &   3.23$\pm$1.05     &1.47$\pm$ 0.08 &		     52$\pm$14~ 	  &0.8      & NON-R\,     \\
  020217890  &1.5129$^{\rm \,a}$&H$\alpha$             &$ 10.02^{+0.16}_{-0.16}$  & $\dots$    & $\dots$                &$\dots$&   $\dots$           &3.57$\pm$0.14 &		     $\dots$ 		  &$\dots$  &$\dots$~     \\
  020239133  &1.0194           &H$\alpha$             &$  9.89^{+0.15}_{-0.15}$  & $9.49$     & $10.59^{+0.18}_{-0.17}$  &   4.8 &   3.09$\pm$0.08     &3.14$\pm$ 0.09 &		     148$\pm$33$^{\rm \,d}$  &2.0      & ROT\,	  \\
  020240675  &1.3270           &H$\alpha$             &$  9.96^{+0.18}_{-0.18}$  & $9.50$     & $10.17^{+0.62}_{-0.62}$  &   3.8 &   2.85$\pm$0.75     &1.09$\pm$ 0.16 &		     47$\pm$92$^{\rm \,d}$	  &1.4      & NON-R\,     \\
  020255799  &1.0351           &H$\alpha$$^{\rm c}$     &$  9.87^{+0.16}_{-0.16}$  & $9.44$     & $10.37^{+0.36}_{-0.36}$  &   4.0 &   2.91$\pm$0.08     &1.89$\pm$ 0.26 &		     14$\pm$25~	  	  &0.2      & NON-R\,     \\
  020261328  &1.5290           &H$\alpha$             &$ 10.01^{+0.20}_{-0.20}$  & $9.59$     & $10.65^{+0.34}_{-0.28}$  &   5.1 &   3.22$\pm$0.08     &1.83$\pm$ 0.27 &		     120$\pm$33$^{\rm \,d}$  &2.2      & ROT\,	  \\
  020278667  &1.0516           &H$\alpha$$^{\rm c}$     &$ 10.28^{+0.16}_{-0.16}$  & $9.04$     & $ 9.28^{+0.36}_{-0.19}$  &   1.3 &   2.35$\pm$0.08     &2.82$\pm$ 1.66 &		     76$\pm$187$^{\rm \,d}$  &1.5      & NON-R\,     \\
  020283083  &1.2818           &H$\alpha$             &$ 10.05^{+0.21}_{-0.21}$  & $9.52$     & $ 9.89^{+0.23}_{-0.21}$  &   5.6 &   2.76$\pm$0.16     &4.29$\pm$ 0.20 &		     59$\pm$12~	          &1.5      & NON-R\,     \\
  020283830  &1.3949           &H$\alpha$             &$ 10.37^{+0.17}_{-0.17}$  & $9.94$     & $10.81^{+0.03}_{-0.01}$  &   7.9 &   3.96$\pm$0.33     &6.82$\pm$ 0.16 &		     186$\pm$30~	  &11.3     & ROT\,    	  \\
  020294045  &1.0028           &H$\alpha$             &$  9.80^{+0.15}_{-0.15}$  & $9.55$     & $10.93^{+0.21}_{-0.16}$  &   5.5 &   3.85$\pm$0.16     &2.89$\pm$ 0.10 &		     233$\pm$51$^{\rm \,d}$  &3.9      & NON-R\,     \\
  020306817  &1.2225$^{\rm \,a}$  &[O${\rm III}$]        &$  9.76^{+0.20}_{-0.20}$  & $\dots$    & $\dots$             &$\dots$&   $\dots$           &4.22$\pm$ 0.09 &		     $\dots$              &$\dots$  & $\dots$~    \\
  020363717  &1.3339           &H$\alpha$             &$  9.68^{+0.35}_{-0.20}$  & $9.90$     & $11.89^{+0.28}_{-0.28}$  &   6.0 &   3.11$\pm$0.08     &0.72$\pm$ 0.09 &		     43$\pm$85$^{\rm \,d}$	  &0.5      & NON-R\,     \\
  020370467  &1.3338           &H$\alpha$             &$ 10.57^{+0.14}_{-0.14}$  & $10.02$    & $11.22^{+0.00}_{-0.54}$  &   5.5 &   3.53$\pm$0.08     &1.31$\pm$ 0.25 &		     51$\pm$65~	  	  &0.6      & NON-R\,     \\
  020386743  &1.0487           &H$\alpha$             &$  9.88^{+0.20}_{-0.20}$  & $9.80$     & $10.21^{+0.25}_{-0.33}$  &   5.4 &   2.99$\pm$3.32     &2.68$\pm$ 0.21 &		     41$\pm$10$^{\rm \,d}$   &0.8      & NON-R\, 	  \\
  020461235  &1.0349           &H$\alpha$             &$ 10.36^{+0.15}_{-0.15}$  & $9.40$     &  $9.99^{+0.16}_{-0.10}$  &   5.4 &   3.47$\pm$0.16     &3.96$\pm$ 0.12 &		     82$\pm$16~	          &3.5      & ROT\,	  \\
  020461893  &1.0486           &H$\alpha$             &$  9.66^{+0.21}_{-0.21}$  & $9.54$     & $10.65^{+0.23}_{-0.31}$  &   6.5 &   3.56$\pm$0.08     &2.66$\pm$ 0.11 &		     58$\pm$13~	          &0.9      & ROT$^{\rm \,e}$\\
  020465775  &1.3583           &H$\alpha$             &$ 10.12^{+0.20}_{-0.20}$  & $9.99$     & $10.25^{+0.21}_{-0.25}$  &   4.9 &   3.70$\pm$0.08     &4.04$\pm$ 0.16 &		     68$\pm$15~	          &0.8      & NON-R\, 	  \\
  140083410  &0.9435           &H$\alpha$             &$ 10.07^{+0.18}_{-0.18}$  & $9.78$     & $10.62^{+0.35}_{-0.35}$  &   5.3 &   3.00$\pm$0.08     &1.93$\pm$ 0.19 &		     30$\pm$33~	  	  &0.4      & NON-R\,     \\
  140096645  &0.9655           &H$\alpha$             &$ 10.40^{+0.24}_{-0.23}$  & $10.12$    & $11.11^{+0.63}_{-0.63}$  &   4.5 &   3.41$\pm$0.08     &1.80$\pm$ 0.40 &		     295$\pm$710~         &3.9      & ROT\, 	  \\
  140123568  &1.0012           &H$\alpha$$^{\rm c}$     &$  9.73^{+0.39}_{-0.39}$  & $9.24$     &  $8.75^{+0.49}_{-0.49}$  &   0.5 &   1.44$\pm$0.08     &1.39$\pm$ 0.98 &		     50$\pm$97$^{\rm \,d}$	  &0.7 	    & NON-R\,     \\
  140137235  &1.0445           &H$\alpha$$^{\rm c}$     &$ 10.07^{+0.29}_{-0.29}$  & $9.77$     &  $8.78^{+0.31}_{-0.10}$  &   0.7 &   1.54$\pm$1.62     &5.45$\pm$ 2.03 &		     61$\pm$11$^{\rm \,d}$	  &3.5      & ROT\,       \\
  140217425  &0.9792           &H$\alpha$             &$ 10.84^{+0.17}_{-0.17}$  & $10.45$    & $11.85^{+0.04}_{-0.02}$  &  14.5 &   5.50$\pm$0.80     &8.96$\pm$ 0.13 &		     320$\pm$46$^{\rm \,d}$  &7.1      & ROT\,   	  \\
  140258511  &1.2423           &H$\alpha$             &$ 11.82^{+0.06}_{-0.09}$  & $10.21$    & $10.41^{+0.33}_{-0.33}$  &   5.2 &   3.17$\pm$0.08     &2.62$\pm$ 2.11 &		     124$\pm$26~	  &5.1      & ROT\,       \\
  140262766  &1.2836           &H$\alpha$             &$  9.84^{+0.43}_{-0.43}$  & $9.38$     & $10.39^{+0.30}_{-0.30}$  &   4.1 &   3.09$\pm$0.08     &2.07$\pm$ 0.70 &		     114$\pm$226$^{\rm \,d}$ &3.0      & ROT\,       \\
  140545062  &1.0408           &H$\alpha$             &$ 10.60^{+0.18}_{-0.18}$  & $9.61$     & $11.10^{+0.40}_{-0.40}$  &   7.5 &   2.93$\pm$0.24     &2.89$\pm$ 6.86 &		     202$\pm$45$^{\rm \,d}$  &3.0      & ROT\,    	  \\
  220014252  &1.3105           &H$\alpha$             &$ 10.78^{+0.21}_{-0.21}$  & $10.42$    & $11.35^{+0.21}_{-0.26}$  &  10.3 &   4.95$\pm$0.16     &3.32$\pm$ 0.12 &		     129$\pm$27~	  &1.4      & ROT\,       \\
  220015726  &1.2933           &H$\alpha$             &$ 10.77^{+0.27}_{-0.27}$  & $10.09$    & $10.72^{+0.41}_{-0.20}$  &   3.7 &   2.85$\pm$0.08     &2.76$\pm$ 0.25 &		     231$\pm$354~         &3.7      & ROT\,       \\
  220071601  &1.3538$^{\rm \,a,b}$ &H$\alpha$             &$ 10.81^{+0.62}_{-0.56}$  & $\dots$    & $\dots$             &$\dots$&   $\dots$	      &8.13$\pm$ 0.13 &		     $\dots$              &$\dots$  & $\dots$~    \\
  220148046  &2.2442$^{\rm \,b}$  &[O${\rm III}$]$^{\rm c}$ &$ 11.22^{+0.17}_{-0.17}$ & $10.10$    &  $8.72^{+0.06}_{-0.06}$  &   0.9 &   0.82$\pm$0.08     &1.98$\pm$ 0.38 &		     61$\pm$119~          &1.3      & NON-R\,     \\
  220376206  &1.2445           &H$\alpha$             &$ 10.67^{+0.27}_{-0.26}$  & $10.51$    & $11.14^{+0.12}_{-0.11}$  &  10.0 &   5.42$\pm$0.08     &5.29$\pm$ 0.08 &		     201$\pm$27~	  &2.8      & ROT\,  	  \\
  220386469  &1.0226$^{\rm \,b}$  &H$\alpha$$^{\rm c}$      &$ 10.80^{+0.16}_{-0.16}$  & $9.64$    &  $9.25^{+0.31}_{-0.32}$  &   2.6 &   1.77$\pm$0.08     &2.90$\pm$ 0.13 &		     98$\pm$20$^{\rm \,d}$	  &2.3      & NON-R\,     \\
  220397579  &1.0379           &H$\alpha$             &$ 10.23^{+0.17}_{-0.17}$  & $10.23$    & $10.97^{+0.23}_{-0.33}$  &  10.2 &   3.00$\pm$0.07     &3.02$\pm$ 0.32 &		     9$\pm$10$^{\rm \,d}$    &0.2      & NON-R\, 	  \\
  220544103  &1.3973           &H$\alpha$             &$ 10.71^{+0.27}_{-0.27}$  & $10.27$    & $10.70^{+0.00}_{-0.20}$  &   7.6 &   5.31$\pm$0.08     &5.61$\pm$ 19.8 &		     137$\pm$24~	  &1.9      & ROT\,   	  \\
  220544394  &1.0101           &H$\alpha$             &$ 10.34^{+0.23}_{-0.23}$  & $9.97$     &  $9.98^{+0.21}_{-0.24}$  &   5.0 &   4.49$\pm$0.08     &3.40$\pm$ 0.15 &		     55$\pm$11~	          &1.1      & NON-R\,	  \\
  220576226  &1.0217           &H$\alpha$             &$ 10.31^{+0.23}_{-0.23}$  & $9.97$     & $10.48^{+0.28}_{-0.28}$  &   6.1 &   3.22$\pm$0.08     &2.16$\pm$ 0.12 &		     30$\pm$12~	          &0.6      & NON-R\, 	  \\
  220578040  &1.0462           &H$\alpha$             &$ 10.72^{+0.17}_{-0.16}$  & $9.80$     & $11.05^{+0.41}_{-0.22}$  &   7.0 &   6.55$\pm$0.08     &3.81$\pm$ 0.12 &		     244$\pm$211$^{\rm \,d}$ &4.9      & ROT\,       \\
  220584167  &1.4655           &H$\alpha$             &$ 11.21^{+0.24}_{-0.24}$  & $10.51$    & $11.28^{+0.07}_{-0.05}$  &  13.1 &   5.24$\pm$0.08     &7.17$\pm$ 0.19 &		     235$\pm$35~	  &4.8      & ROT\,   	  \\
  220596913  &1.2658$^{\rm \,b}$  &H$\alpha$             &$ 10.68^{+0.30}_{-0.30}$  & $9.53$     & $10.66^{+0.04}_{-0.04}$  &   9.3 &   5.09$\pm$0.08     &9.49$\pm$ 11.5 &		     141$\pm$10~	  &3.7      & ROT\,   	  \\
  910193711  &1.5564$^{\rm \,b}$  &H$\alpha$             &$  9.99^{+0.42}_{-0.18}$  & $10.14$    & $10.39^{+0.31}_{-0.44}$  &   4.1 &   1.61$\pm$0.08     &2.27$\pm$ 0.06 &		     63$\pm$12~	          &0.8      & NON-R\,	  \\
  910279515  &1.4013$^{\rm \,b}$  &H$\alpha$$^{\rm c}$     &$ 10.79^{+0.14}_{-0.14}$  & $9.71$     & $10.34^{+0.07}_{-0.05}$  &   2.7 &   4.05$\pm$0.25     &4.44$\pm$ 0.11 &		     265$\pm$18~          &5.6      & ROT\,   	  \\ [0.3mm]
\hline \hline                                                                                                                                                                             
\end{tabular}                                                                                                                                                                       
\end{center}                                                                                                                                                                       
\noindent                         
\label{tab:log}
{{\bf Tab.\,2.} 
  (1) VVDS identification number.
  (2) Redshift derived from SINFONI observations of the emission line 
(\halpha$\lambda$6563\AA\ or \oiii$\lambda$5007\AA) tabulated in column (3).
  (4) Stellar mass in solar mass units.
  (5) Gas mass derived using the Kennicutt-Schmidt formulation \citep{kennicuttschmidt98}.
  (6) Dynamical mass computed following Eq.~1 and
  based on the best-fitting parameters of the kinematic model.
  {(7) The radius computed as the total size of the ionized emission map with a confidence
level larger than $2\sigma$.
  (8) The gas half-light radius estimated on the SINFONI maps of the lines tabulated in column (3).
  (9) The observed I-band half-light radius.}
  (10) Maximum velocity deduced from model fitting.
  (11) Ratio of the maximum rotation velocity over the velocity dispersion.
  (12) Kinematic classification (R: Rotating objects, NON-R: non-rotating systems).
  $^{\rm a}$~This redshift is computed on the original VIMOS spectrum.
  $^{\rm b}$~Observations with Adaptive Optics.
  $^{\rm c}$~Galaxy with low signal-to-noise ratio ($3\le S/N\le4.5$) plotted with empty
symbols in the figures. ~Galactic properties of SINFONI observations with a 
signal-to-noise ratio below 3 are not reported. 
  $^{\rm d}$~Galaxy for which the plateau velocity has not been reached.
  $^{\rm e}$~Galaxy for which the criterion on $v/\sigma>1$ is not fulfilled, thus not included in the purely rotating disc sample.
   {The public access to data is at http://cosmosdb.iasf-milano.inaf.it/VVDS-SINFONI.}}
\end{table*}                       

\section{Derived quantities}

 \subsection{Morphology and kinematics}
\label{sec:fitting}

We used the \halpha$\lambda$6563\AA\ line (and \oiii$\lambda$5007\AA\ line in
one case) to derive the dynamical properties. We assumed that the
ionized gas rotates in a thin disc with a velocity reaching a plateau
in the outer regions (but for 8/23 rotating galaxies marked with
$^{(c)}$ in column (9) of Tab.\,2, this radius was not reached within
the area covered by our observations). Using a $\chi^{2}$
minimization, we produce seeing-corrected velocity and dispersion maps
of galaxies with geometrical inputs weighted by the S/N of each pixel.

The input geometrical parameters of the fitting model were estimated
from the I-band best-seeing CFHTLS\footnote{Based on observations
  obtained with MegaPrime/MegaCam, a joint project of CFHT and
  CEA/DAPNIA, at the Canada-France-Hawaii Telescope (CFHT) which is
  operated by the National Research Council (NRC) of Canada, the
  Institut National des Science de l'Univers of the Centre National de
  la Recherche Scientifique (CNRS) of France, and the University of
  Hawaii. This work is based in part on data products produced at
  TERAPIX and the Canadian Astronomy Data Centre as part of the
  Canada-France-Hawaii~~Telescope Legacy Survey, a collaborative
  project of NRC and CNRS.} Megacam images for all galaxies, apart
from 14$^{\rm hr}$ galaxies that were covered with the I-band CFHT-12K
camera only \citep{maccraken03}. At the typical $z\sim 1.2$ redshift
of {\small MASSIV} galaxies, I-band images probe the
3200-4200\AA\ rest-frame wavelength range, or the U-band. We used the
GALFIT software \citep{peng02} that convolves a PSF with a model
galaxy image based on the initial fitting parameter estimates
determined by fitting a S\'ersic (1968) profile. GALFIT converges into
a final set of parameters such as the centre, the position angle, and
the axial ratio. Residual maps from the fit were used to optimize the
results. The I-band images were also used to calibrate for SINFONI
astrometry, by considering the relative positions of the PSF star.
GALFIT was also applied to the ionised gas maps (properly deconvolved
with a PSF star) to derive the semi-major axis disc scale-length. The
CFHT images and the kinematic maps are presented in Epinat et
al. (2012), where more discussion is presented on the modelling
procedure.

The major source of uncertainties in deriving the maximum rotation
velocity is the inclination of the studied galaxy. For very small
objects, the inclination is not robustly constrained, thus for a
fraction of galaxies the median inclination value of a random
distribution of galaxies on the sky (60\degr) is adopted instead. For
this category of objects, we estimated that a typical uncertainty of
24\degr\ is associated with the inclination. Another uncertainty appears
when the maximum velocity is not reached within the radius covered by
our observations. Other uncertainties entering into the error budget
are related to the simplistic disc model adopted. The evaluation of the net
effects of this assumption would require some assumption about the different,
detailed modelling schemes, which is beyond the scope of this
paper. The final uncertainty associated with our velocities includes
both the errors originating from the deprojection of the radial
velocity given a certain inclination (estimated with Monte Carlo
simulations) and the errors in the modelling procedure (quantified using
simulations of the GHASP local sample following \citealt{epinat11}).

\subsection{Mass and star formation rate}
\label{sec:mass}

We inferred the stellar mass from a spectral energy distribution (SED)
fit to photometric and spectroscopic data with BC03 stellar population
synthesis models \citep{bc03} using the GOSSIP spectral energy
distribution tool \citep{fra08}. We assumed a Salpeter initial mass
function \citep{sal55} with a lower and upper mass cutoff of
respectively 0.1 and 100\msun, and a set of delayed exponential star
formation histories with galaxy ages in the range from 0.1 to
15~Gyr. As input to the SED fitting, we used the multi-band
photometric observations available in the VVDS fields, including BVRI
data from the CFHT-12K camera, ugriz data from the CFHT Legacy Survey,
J and Ks-band data from SOFI at the NTT, from the UKIDSS survey, SWIRE
data when available, and the VVDS spectra. Following
\citet{walcher08}, we adopted the probability distribution function to
obtain the stellar mass (listed in column (6) of Tab.\,2), the
absolute magnitude, and other results of the fitting procedure as
detailed in \citet{contini11}.

We estimated the dynamical mass and the gas mass in our galaxies to
trace the evolution of both the baryonic and dark matter
components. As described in the next section our galaxies have a
variety of kinematical properties that we classify into two broad
categories: rotating and non-rotating systems (see details in
Sect.\,3.3 of this paper and \citealt{epinat11}). The dynamical mass
is estimated using the equation

\begin{equation}
M_{dyn} = \frac{V^2~R}{G} + \frac{\sigma^2~R^3}{G~h^2}\,,
\end{equation}

\noindent where the first term represents the mass enclosed within the
radius $R$ defined as the total extent of the ionized gas, $V$ is the
velocity at the same radius, and $G$ is the universal gravitational
constant. The second term represents the asymmetric drift correction
following \citet{meurer96}, where $\sigma$ is the dispersion velocity
at $R$, and $h$ is the gas disc scale length in an isotropic
geometrical configuration with a surface density described by a
Gaussian function. The dynamical mass of each galaxy is given in
column (6) of Tab.\,2, where its error was estimated by accounting for
the uncertainties related to the inclination and the modelling
procedure.

The gas mass was derived using the empirical correlation between star
formation rate (SFR) and gas surface density assuming that the gas and
the stellar content accounts for the majority of the mass in the
central regions of galaxies. In the local Universe,
\citet{kennicuttschmidt98} proved that the conversion of the surface
density of gas to star-formation rate follows a power law above some
critical gas density \citep[$\Sigma_{SFR} =
 A\Sigma_{gas}^n$]{schmidt59} over more than six orders of magnitude
in $\Sigma_{\rm SFR}$. This empirical correlation has not yet been
routinely proven to hold at high redshift because of the lack of
extensive measurements of cold gas mass, but a number of studies of
small samples support the validity of this relation at high redshifts
\citep[e.g.,][]{erb2006, bouche07, daddi2010, tacconi2010}. Some
detailed studies of individual galaxies also highlight a consistency
with the local Schmidt law (e.g., the lensed $z=2.7$ LBG MS1512-cB58,
\citealt{baker2004}). Thus, we computed the star-formation rate
surface densities using the total SFR and the size of our galaxies.

The SFRs were computed following \citet{kennicutt98}. The fluxes were
measured in $>2\sigma$ regions of the flux-calibrated data cubes and
corrected for dust reddening using the extinction coefficient derived
from the SED fitting \citep[see][]{contini11}. The half-light radius,
$R_{1/2,gas}$, was measured with GALFIT \citep{peng02} as the
semi-major-axis half-light radius describing the spatial extent of the
ionized gas (\halpha$\lambda$6563\AA\ or \oiii$\lambda$5007\AA)
emission, after convolving the images with the PSF of a bright star
nearest in time to the scientific observations. Maps of the residuals
were used to optimize the results. Assuming that the Schmidt law holds
in our redshift interval, we estimated the gas mass ${\rm M_{\rm
      gas}}$ using the relation between the gas density $\Sigma_{\rm
    gas}$ and the star-formation rate surface density $\Sigma_{\rm
    SFR}$ \citep{kennicuttschmidt98}

\begin{equation}
  \frac 
  {\Sigma_{\rm SFR}}
  {\rm M_{\odot}\; yr^{-1}\; kpc^{-2}}
  = 2.5 \times 10^{-4} 
  \left(
  \frac 
  {\Sigma_{\rm gas}}
  {\rm M_{\odot} pc^{-2}} 
  \right)^{1.4}\,,
\end{equation}

\noindent where the relation between the observed \halpha\ luminosity per unit
area, or the \halpha\ surface density $\Sigma_{\rm H\alpha}$, and the
gas surface density is

\begin{equation}
  \frac
  {\Sigma_{\rm gas}}
  {\rm    M_{\odot}\; pc^{-2}}
  =1.6\times10^{-27}
  \left( 
    \frac{\Sigma_{\rm H\alpha}}{{\rm erg\; s^{-1}\; kpc}^{-2}}
  \right)^{0.71} \,.
\end{equation}

\noindent The mass of the gas associated with the measured star formation is given by

\begin{equation}
 M_{\rm gas}=\Sigma_{\rm gas}\times {\it Area}\,,
\end{equation}

\noindent where the {\it Area} in Eq.~4 is the total spatial extent of
the ionized gas measured in $>2\sigma$ regions of the flux-calibrated
data cubes.

The gaseous masses are given in Tab.\,2, and their associated error
was analytically propagated from the errors in the observations using
conventional techniques. Further details of the error analysis are
described in \citet{epinat11}. Typical value of the error is of the
order of $0.25$~dex. We note that different calibrations exist for the
Kennicutt-Schmidt law at higher redshift, e.g., \cite{bouche07}
suggest a steeper power-law index (1.7 instead of 1.4 in Eq.\,3) for
$z>2.5$ starburst galaxies. This steeper formulation implies a $<
0.3$~dex difference in the gas computation. Given the uncertainties
involved in deriving this quantity, a different value for the slope
does not have any significant impact on our data interpretations (see
Gnerucci et al. 2011).

\subsection{Kinematic classification}
\label{sec:class}

To classify our sample, we adopt a kinematic classification that relies
on a combination of observed galaxy properties. We refer to
\citet{epinat11} for a full description of the methodology of which
we provide here a brief outline. In a first step, eight of us
classified independently all the sample using personal
criteria. Subsequently, we reconciled the criteria into a newly agreed
scheme: we assigned to each galaxy a confidence flag generated by the
concordance among the classifiers.

When galaxies satisfy all the following conditions, they are
kinematically flagged either as:
\begin{itemize}
\item[]{\bf Rotating discs:} {when the velocity field is well
  described by a symmetrically rotating disc with a kinematic position
  angle that does not differ significantly from the morphologically
  derived major axis. We quantify these conditions assuming that the
  mismatch between the position angles is smaller than
  $\pm$20\degr\ and that the residuals of the modelled velocity field
  over the measured velocities are less than a certain fraction
  ($<20$\%). In this work, galaxies are classified as ``rotators''
  when they are rotationally supported such that their rotation
  velocity is larger than their dispersion velocity (i.e.,
  $v/\sigma>1$ as listed in column (12) of Tab.\,2)},
\end{itemize}

\noindent or

\begin{itemize}
\item[]{\bf Non-rotating systems:} {At least one of the conditions about
  the residuals, position angles, or rotation-over-chaotic motions is
  violated. This class includes galaxies with some kind of anomalies in
  their velocity field, merger-like systems, and/or objects with
  off-peak distributions of the dispersion velocity, but also
  possibly slow rotators or face-on systems with an observed velocity
  gradient smaller than typically $\pm25$\kms.}
\end{itemize}

\section{The gas properties in {\small MASSIV} star-forming galaxies}

\begin{figure}
\begin{center}
\includegraphics[width=9.5cm,angle=0]{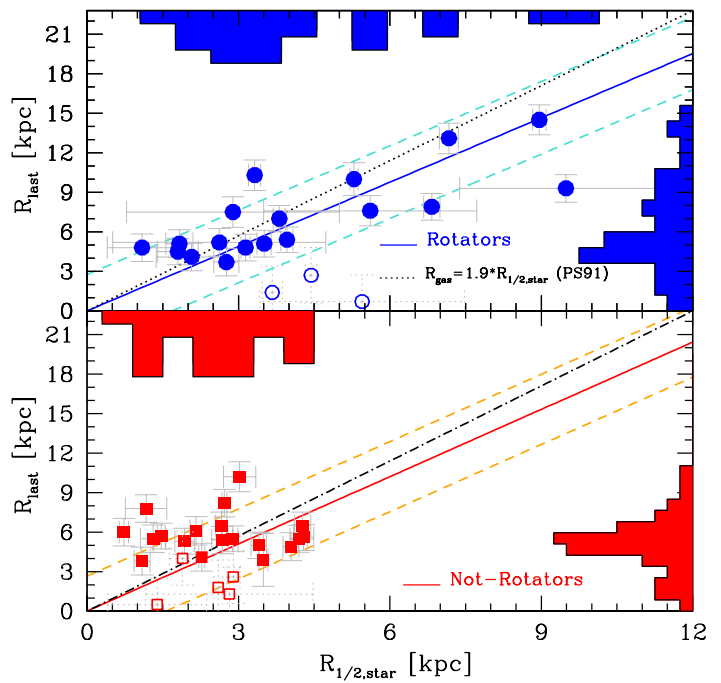}
\end{center}
\caption{Comparison between ${\sl R}_{last}$ and $R_{1/2,star}$ radii
  computed as the size of our ionized gas maps with a confidence level
  larger than $2\sigma$ and the half-light radius of the stellar
  continuum measured on the observed I-band best-seeing CFHTLS images,
  respectively. Within the errors, both rotators (top panel, blue
  circle) and non-rotators (bottom panel, red square) agree with the
  expected statistical trend $R_{last}=1.9 \times R_{1/2,star}$ found
  by Persic \& Salucci (1991, PS91) plotted with a black dotted line.
  The blue/red solid lines are the best-fit relation found for the
  rotators/non-rotating galaxies (dashed lines show the total scatter
  of the correlations). The empty symbols show galaxies
  detected with a low S/N ($3\le S/N\le4.5$). \label{fig:radii} }
\end{figure}

We compare the total extent of the ionized gas, ${\sl R}_{last}$, with
the half-light radius of the stellar component, $R_{1/2,star}$, for
the class of rotating and non-rotating disc galaxies in
Fig.\,\ref{fig:radii} (in the top panel with blue colour-coded circles
and in the bottom panel with red colour-coded squares,
respectively). The radius ${\sl R}_{last}$ is computed as the total
size of the ionized emission map that has a confidence level larger
than $2\sigma$. The half-light radius of stars is measured on the
observed I-band images, or at rest-frame U-band wavelengths. The error
bars in Fig.\,\ref{fig:radii} are the $1\sigma$ uncertainties derived
from the GALFIT procedure on the images corrected for their respective
PSF FWHM. The histograms show the distributions of the gas and stellar
radii of the classes. The solid line is the linear fit\footnote{We fit
  a linear relationship to the data using the MPFITEXY routines
  \citep{williams2010}. The MPFITEXY routine depends on the MPFIT
  package \citep{markwardt2009}. This routine properly adjusts the
  intrinsic scatter to ensure the minimisation of chi-square with an
  iterative prescription.} to the data and the dashed line shows the
standard deviation.

The dotted line is the correlation observed in a thin exponential disc
between the radius of the total ionized gas emission and the
half-light radius of the stellar component, $R_{last} \sim 1.9 \times
R_{1/2, star}$. Assuming that the optical radius equals 3.2 times the
disc scalelength following \citet{persic91}, the half-light radius is
1.678 times the disc scale length. The total ionized gas, $R_{last}$,
is close to the optical radius measured for local star-forming
galaxies \citep[e.g.,][]{garrido2005}.

For the class of rotators, we obtained median values of ${\sl
  R}_{last}=5.40^{7.60}_{4.80}$~kpc and
$R_{1/2,star}=3.51^{5.29}_{2.62}$~kpc, and for non-rotators ${\sl
  R}_{last}=5.50^{6.50}_{4.00}$~kpc and
$R_{1/2,star}=2.68^{3.40}_{1.47}$~kpc (with the extremes representing
the quartiles of the distribution for the two classes). Non-rotating
galaxies were found to have more compact stellar components than
rotators, but similarly sized gas components. For rotating galaxies,
we found that $R_{last}= 1.63\pm0.14 \times R_{1/2,star}$ with an
intrinsic scatter of $\sigma_{intr}=2.23$ (and total scatter of
$\sigma_{tot}=2.70$). The same correlation for non-rotating systems is
$R_{last}= 1.70\pm0.21 \times R_{1/2,star}$ ($\sigma_{intr}=2.20$ and
$\sigma_{tot}=2.66$). In the local Universe, the size of the ionized
gas component of star-forming galaxies is close to the optical radius
\citep[e.g.,][]{garrido2005}. Within 1$\sigma$ dispersion, and
considering the associated uncertainties, all {\small MASSIV} rotating
galaxies are within the expected statistical trend found by
\citet{persic91}. A similar agreement was found by Puech et al. (2010)
for their IMAGES sample at $z\sim 0.6$. As for Puech et al. (2010) and
\citet{banford2007}, we note that the fraction of non-rotating
galaxies with a larger gaseous extent than their stellar counterpart
exceeds the predicted correlation by $1\sigma$. While in the sample of
Puech et al. (2010) the cases where this correlation is not respected
are the ones with the UV light extending further out the field-of-view
of the FLAMES/IFU ($3\arcsec \times2\arcsec$), in the {\small MASSIV}
sample the two outliers show a signature of a close companion, one of
which is detected in \halpha.

We will investigate these trends and their relation to the
environment at the completion of the survey when we will develop a
classification scheme enabling us to distinguish the various
mechanisms of disturbance in inclined and face-on galaxies as well as
in compact and dispersion-dominated spheroids. For simplicity, at the
present time we flag all these galaxies in a unique class named
non-rotators (plotted with red-coded, square symbols in all figures).

We now compare our measurements with the characteristic sizes of
galaxies measured by other similar surveys to analyse the correlation
among properties such as stellar mass, size, and velocity. In IMAGES
(Puech et al. 2010), the total light radius measured in observations
of the ionized [OII] gas of rotating galaxies has a median of
11.29~kpc and ranges from 8.25~kpc (first quartile, or 25th percentile
of distribution) to 11.99~kpc (last quartile, or 75th percentile). The
median value of the half-light rest-frame UV radius is 6.27~kpc where
4.26~kpc and 6.53~kpc are quartiles of the distribution. We note that
the IMAGES disc size reaches a maximum between the total size measured
in the ionized [OII] gas and the half-light rest-frame UV light
multiplied by 1.9. In SINS (Cresci et al. 2009), the half-width
half-maximum size (HWHM) measured on the ionized \halpha\ maps by
\citet{bouche07} is interpreted as the exponential disc scale-lengths
of $z\sim2$ rotating SINS discs, and ranges from 4.11~kpc to 6.35~kpc
with a median of 5.87~kpc. The circular half-light sizes of the total
SINS $z>2$ sample computed on \halpha\ maps by \citet{nfs09} is in the
range $2.40-4.60$~kpc and has a median value of 3.10~kpc.
\citet{dutton2011} claim that there is a fundamental disagreement
between the SINS HWHM sizes published in Cresci et al. (2009) and the
SINS half-light radii measured by \citet{nfs09}, as already noted by
these latter authors who reject the possibility that this discrepancy
is caused by non-exponential discs. At higher redshift, the AMAZE/LSD
sample of \citet{gnerucci2011} is constituted of eleven $z\sim3$
rotating galaxies. The characteristic radius of these exponential
discs measured in \halpha\ maps ranges between 0.71~kpc and 1.73~kpc
(first and last quartile of the distribution) with a median value of
1.24~kpc. Comparing directly to the values of Puech et al. (2010)
since we adopt the same radius definition, we find overall that IMAGES
radii are a factor of 1.5 larger than our {\small MASSIV} radii. When
comparing with the radii derived for SINS by \citet{nfs09}, we find
similar sizes (our half-light sizes measured in our \halpha\ maps
being in the interval $R_{1/2,gas} =2.85-3.96$~kpc with a median of
3.47~kpc). Both SINS and {\small MASSIV} radii are approximately three
times larger than sizes sampled at $z\sim 3$ in AMAZE/LSD by Gnerucci
et al. (2011).

We investigate the disc properties to verify the hypothesis of whether
our {\small MASSIV} discs have exponential profiles. Discs of ionized
gas frequently have an irregular, clumpy distribution, as previously
reported for SINS galaxies. Despite the simplistic assumptions, both a
S\'ersic index distribution and a correlation between exponential and
half-light radii suggest that {\small MASSIV} discs can be
approximated as exponential discs. The same qualitative conclusion was
reported for SINS discs \citep{nfs09}.

In summary, our resulting characteristic size agrees within the
statistical correlations previously assessed in the literature. The
radius taken to derive the velocity is critical to avoid the
introduction of systematic biases \citep[see][]{noordmermeer07}. The
relation obtained between the size of the stars and that of the
gaseous component ensures that this latter physical quantity is
reliable for the determination of the maximum velocity.

\begin{figure}[ht!]
\begin{center}
\includegraphics[width=8cm,angle=0]{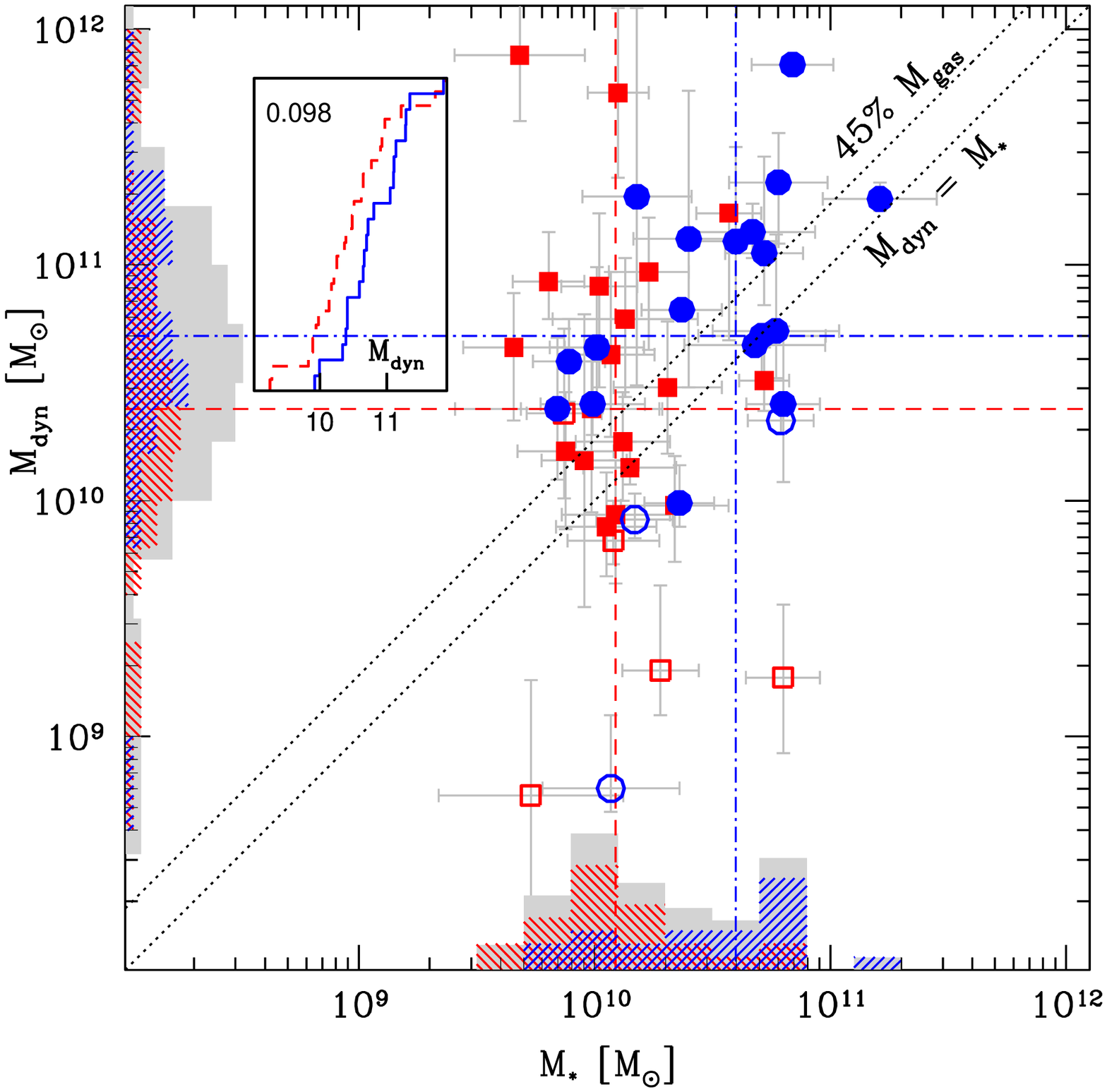} 
\includegraphics[width=8cm,angle=0]{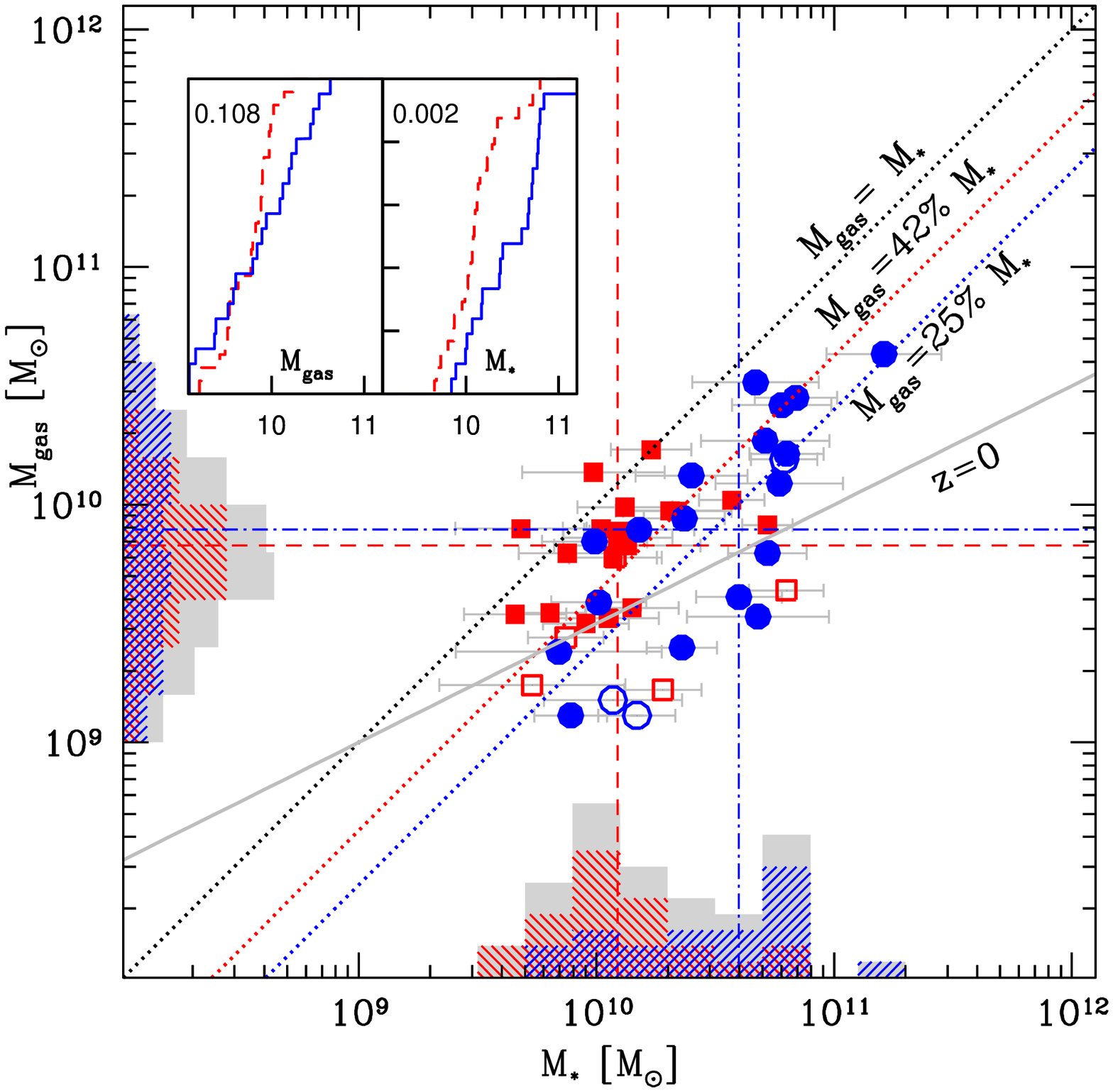} 
\end{center}
\caption{The stellar mass content of {\small MASSIV} galaxies compared
  to both the dynamical mass derived from the dynamical modelling (top
  panel) and gas content derived using the Kennicutt-Schmidt
  formulation (bottom panel). Symbols are as in Fig.\,1. The
  horizontal and vertical histograms represent the distribution of the
  stellar and dynamical masses (top) and gas mass (bottom panel). The
  horizontal and vertical lines are the median values of rotators
  (point-dashed blue line) and non-rotators (dashed red line),
  respectively. Dashed diagonal lines in the bottom panel comply with
  the condition $M_{gas}= frac \times M_{star}$, where $frac$ is equal
  to 25\% for rotating galaxies and 42\% for non-rotators (or 20\% and
  30\% of the baryonic mass, respectively). The solid line is the
  local relation proposed by \cite{schim08} using the GALEX Arecibo
  SDSS survey of cold \hi\ gas. In the inserted panels, the
  cumulative distributions for stellar-, dynamical-, and gas-mass are
  plotted for rotators (solid, blue line) and non-rotators (dashed,
  red line) along with the probability for the two classes to be
  statistically different. \label{fig:gas} }
\end{figure}

Figure\,\ref{fig:gas} shows the stellar mass of {\small MASSIV}
galaxies versus the dynamical (top panel) and gas (bottom panel) mass.
The methods used to derive these masses are described in
Sect.\,\ref{sec:mass}. The total distributions of these quantities are
shown as full grey-coded histograms, while those with blue,
positive-oblique angle (or red negative-oblique angle symbols) refer
to rotating (non-rotating) galaxies. Dynamical masses range between
$2.57\times 10^{10}$~\msun\ and $1.29 \times 10^{11}$~\msun\ with a
median value of $5.01\times 10^{10}$~\msun\ for the class of rotating
galaxies (dashed-point horizontal line in the bottom panel of
Fig.\,\ref{fig:gas}), and $0.95-8.13 \times 10^{10}$~\msun\ with a
median value of $2.45 \times 10^{10}$~\msun\ for non-rotating galaxies
(dashed horizontal line). We estimated a gaseous mass in the interval
between $3.38\times 10^{9}$~\msun\ (first quartile) and $1.64\times
10^{10}$~\msun\ (last quartile) with a median value of $7.85\times
10^{9}$~\msun, and stellar masses in the range $1.48-5.87 \times
10^{10}$~\msun\ with a median value of $3.98 \times 10^{10}$~\msun\ in
rotating galaxies. Median values for non-rotators are
$M_{star}=1.23\times 10^{10}$~\msun\ and $M_{gas}=6.71\times
10^{9}$~\msun\ with ranges between $\Delta M_{star}=0.90 - 1.90\times
10^{10}$~\msun\ and $\Delta M_{gas}=3.51 - 9.37\times 10^{9}$~\msun.

Within their uncertainties, only a few galaxies lie in the forbidden
region of $M_{star}>M_{dyn}$. The only strongly deviating rotating
galaxy (empty, dashed symbol) was detected with a low S/N ($\sim
3.5$). We inferred a probability of $P = 7.5 \times 10^{-2}$ and $P =
1.5 \times 10^{-4}$ that the stellar mass is uncorrelated to both the
dynamical and gas mass using the Spearman correlation test. The
correlation is statistically significant in particular for the stellar
and gas masses. In the insert panels of Fig.\,\ref{fig:gas}, we show
the cumulative distributions of these quantities for rotators (solid,
blue line) and non-rotators (dashed, red line), along with the
probability that the two classes are statistically different based on
the Kolmogorov-Smirnov test.

We assume no contribution from dark matter in the inner regions of
galaxies. This is the most likely hypothesis given our inability to
measure the dark matter distribution in high$-z$ galaxies. This
hypothesis was adopted in earlier studies, e.g., by Gnerucci et
al. (2011). With this assumption, the mass in gas taken as the
difference between the dynamical and stellar masses of both classes of
rotator and non-rotator galaxies, is on average a fraction of $\sim
45\%$ of the dynamical mass (diagonal dotted line), and approximately
either 20\% of the stellar mass or 17\% of the baryonic mass. A
consistency picture is obtained for rotating galaxies when deriving
the gas mass from the Kennicutt-Schmidt law. We found that the gas
mass is 25\% of the stellar mass (or 20\% of the baryonic mass) in
rotating galaxies with typical errors associated with the fit of the
order of $5-6$\%.

The gas content in non-rotating galaxies is definitively higher ($\sim
42$\% of the stellar mass, or 30\% of the baryonic mass) when using
the Kennicutt-Schmidt law. Taking into account that this class of
galaxies may contain face-on discs, this fraction represents a lower
limit. The same value derived as a difference between dynamical and
stellar mass is somehow lower. It can be justified by two factors: 1)
the inclusion of merging systems in the class of non-rotating galaxies
that may have not yet exhausted their gas reservoir; 2) the different
properties of dark matter halos in rotators and spheroidal
galaxies. In nearby galaxies, within the optical radius the dark
matter content is suspected to be higher in rotators than in spheroids
\citep{persic1996}. Thus, for the same dynamical mass, the fraction of
gas and stars should be lower in rotators than in our class of
non-rotating systems.

If we fix the slope to the local relation proposed by \cite{schim08}
using the GALEX Arecibo SDSS survey of cold \hi\ gas, we obtain a weak
evolution of $+0.17$~dex to $z=1.2$ for the entire sample (or
$+0.11$~dex for rotators and $+0.21$~dex for non-rotating galaxies).
We emphasize that the selection criteria adopted to build up the
{\small MASSIV} sample lead to a subsample of non-rotating galaxies with
a lower content of stellar mass at fixed gas and dynamical mass.

The fraction of gas in $z\sim 0.6$ IMAGES galaxies ranges between 30\%
as derived from the evolution of the gas-metallicity relation
\citep{rodrigues2008} and $45$\% of the stellar mass obtained with the
inverse Kennicutt-Schmidt law by Puech et al. (2010). The gas
fraction quoted by Cresci et al. (2009) for SINS galaxies at $z\sim2$
is in the interval $23-30$\% of the dynamical mass, but including a
dark matter contribution of 40\%. At higher redshift ($z>3$, the
AMAZE/LSD surveys), these values are too uncertain
\citep{gnerucci2011}.

On the basis of direct estimates using molecular observations
\citep[e.g.,][]{daddi2010,tacconi2010}, the gas mass in disc galaxies
has been found to be 34-44\% of the baryonic mass, and have slowly
decreased in the past 8-10~Gyr. Our results are in full agreement with
a slow decrease in the gas mass content in disc galaxies since
$z\sim1.2$ with a gas fraction that then halves progressively prior to
the redshift of the local Universe.

\section{The size - mass - velocity relation}

We now examine the fundamental relations existing in the
{\small MASSIV} sample at $1.0 < z < 1.6$. We compare their sizes,
stellar masses, and rotation velocities to galaxy samples available in
the literature at different redshifts to investigate the evolution of
these properties. In particular, we derive the relation between the
luminosity (or stellar mass) and rotational velocity introduced by
\citet{tf77}, the so-called Tully-Fisher relation (TFR). We explore the
shift in the zero point of this relation and the other fundamental
relations that unveil the inter-connection between dark and luminous
matter for galaxies at different cosmic epochs.

\subsection{The size - mass relation}

\begin{figure}[hb!]
\begin{center}
\includegraphics[width=9.5cm,angle=0]{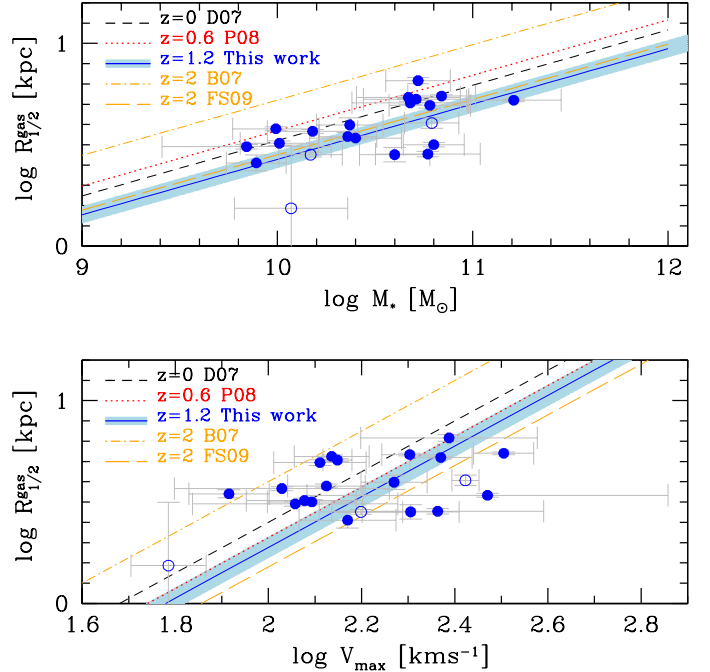}
\end{center}
\caption{The stellar mass$-$size relation ({\bf top panel}) and the
  velocity$-$size relation ({\bf bottom panel}) for {\small MASSIV}
  rotating galaxies at $z\sim1.2$ (solid, blue line with the $1\sigma$
  error of the correlation illustrated by the light blue area). The
  local relation of Dutton et al. (2007, D07) is shown as a
  short-dashed, black line. We overplot the best-fit relations
  obtained by \citet{puech08} (P08, $z\sim 0.6$, dotted red line),
  \citet{bouche07}, and \citet{cresci09} at $z\sim 2$ (B07,
  dotted-dashed orange line), and Dutton et al. (2011) using
  measurements of \citet{nfs09} (FS09, $z\sim 2$, long-dashed orange
  line). The symbols are as in Fig.\,1. \label{fig:velsize}}
\end{figure}

The evolution of the size - stellar mass relation in rotating galaxies
at $1.0 < z < 1.6$ is shown in the top panel of
Fig.\,\ref{fig:velsize}. The distribution of the {\small MASSIV}
rotating galaxies in this plot (blue-coded circles) is consistent with
a mild dependence of the disc size on the stellar mass. Taking into
account the relatively limited statistics associated to a disc size
representing only a lower limit (as measured for galaxies with a S/N
above 3), we interpret this plot in the framework of its evolution
with respect to the local relation by fixing the slope to the local
relation as derived by \citet{dutton2011}. This relation is based on
the local I-band disc scalelength - stellar mass relation of Dutton et
al. (2007) with stellar masses from \citet{bellmci03} and
\citet{bdj01}. The evolution of the relation is only slightly stronger
using the V-band instead of the I-band size-stellar mass relation
derived from the SDSS data. This radius is scaled to the half-light
radius, which is 1.678 times the scale length for the exponential
profile. This local relation is plotted in Fig. 3 as a dashed, black
line. The solid, blue line is the fit to the MASSIV rotating
galaxies. To make a consistent comparison to our measurements, we plot
in this figure only the evolution reported in those studies using the
sizes measured in the ionized gas (\halpha, or \oii) maps. The best
fit obtained using SINS galactic sizes defined as the HWHM H$\alpha$
radii \citep{bouche07,cresci09} is plotted with a dotted-dashed,
orange line and a long-dashed, orange line when the size is defined as
the half-light H$\alpha$ radius by \citet{nfs09}. We note that the
difference in the evolution of this relation is due to the method used
to measure the disc sizes by \citet{bouche07} and \citet{nfs09}
\citep[cf.][]{dutton2011}. We plot with a dotted, red line the
evolution measured for IMAGES rotating galaxies using as disc size the
maximum between the total size measured about the ionized [OII] gas
and the half-light rest-frame UV light multiplied by 1.9. All
quantities are consistently rescaled to our Salpeter IMF.

There is a general consensus about the mild evolution of the disc size in
star-forming galaxies up to $z\sim1$ (for example as measured by
\citet{williams2010} taking the circularized rest-frame I-band
half-light radii of disc-dominated galaxies, and by
\citet{trujillo2006} taking the circularized rest-frame V-band
half-light radius). Compared to the present-day disc sizes, the $z=1$
counterparts are {\it smaller} by $\sim 0.10$~dex at fixed stellar
mass and rotation velocity \citep[e.g.,][]{dutton2011}. This picture
agrees well with the theoretical predictions \citep{blumenthal1984,
  dalcanton1997, mmw1998, firmani2000, dejong2000, pizagno05,
  dutton2007}.

At $z\sim 1.2$, we report an evolution size$-$stellar mass relation of
$-0.14$~dex using the half-light H$\alpha$ radius, that is quite
similar to earlier findings at $z\sim 1$. Above $z\sim 1$, the
evolution of disc sizes in the literature is less clear, even with
opposite trends. Trujillo et al. (2006) found a factor of two smaller
half-light sizes at $z\simeq 2.5$ at fixed stellar mass relative to
$z=0$ galaxies. A similar evolution is reported by Williams et
al. (2010) in star-forming galaxies at $z=2$. The evolution of
circular half-light H$\alpha$ sizes measured by \citet{nfs09} and
reported by Dutton et al. (2011) for SINS data at similar redshifts is
smaller than previously quoted results, i.e., $-0.07\pm0.05$~dex to be
compared with $-0.28$~dex by Trujillo et al. (2006). This discrepancy
is attributed by Dutton et al. (2011) to the difference in the
half-light radius measured on the \halpha\ maps and rest-frame optical
band images. Taking into account this difference, their results can be
reconciled. A different trend is obtained using measurements published
in Cresci et al. (2009) for the same SINS data by Dutton et
al. (2011). Using the half-width half-maximum size interpreted as
exponential disc scale-lengths, SINS disc galaxies presented in Cresci
et al. (2009) are $\sim 1.6$ times {\it larger} at $z=2$ than their
local counterparts at fixed stellar mass. Even if a factor of
$-0.07$~dex caused by the skewed distribution towards low inclinations
reduces the discrepancy, the values remain significantly inconsistent
by a factor of three with the measurements of Trujillo et al. (2006)
and Williams et al. (2010) (cf. Dutton et al. 2011). This discrepancy
between data-sets increases at higher redshifts if we explore the
AMAZE/LSD sample by Gnerucci et al. (2011). The sizes of the discs are
more than four times {\it larger} in the past than in their local
counterparts. We can cautiously speculate that this strong evolution
may be caused by a selection effect in the AMAZE/LSD galaxies and the
difficulties in measuring realistically the \halpha\ sizes at such
high redshifts ($z\sim 3$).

Comparing the evolution computed directly from the half-light radii
measured in H$\alpha$, we found at $z\sim 1.2$ a similar evolution
similar to that derived from SINS data at $z\sim 2$ by Dutton et
al. (2011), based on measurements by \citet{nfs09}. However, if we
account for the shift suggested by Dutton et al. (2011) that converts
our \halpha\ sizes at $z\sim 1.2$ to their optical band half-light
radii, we should have a stronger evolution with redshift, i.e.,
$-0.21$~dex instead of $-0.09\pm0.04$. Conversely, we did not find a
conversion similar to that suggested by Dutton et al. (2011) in the
ratio of galaxy sizes (e.g., of our observed I-band to
\halpha). Furthermore, our results for the evolution of the disc size
perfectly agree with the evolution up to $z\sim 1$ found by
\citet{williams2010} and \citet{trujillo2006}, especially taking into
account the associated uncertainties in the measurements and the
adopted assumptions. Therefore, on the basis of the current data-set
we suggest that there is a mild evolution in the size - stellar mass relation of
rotating galaxies at $z\sim 1.2$. A homogeneous comparison of {\small
 MASSIV} galaxies at $z\sim2$ and $z\sim1$ that will be possible at
the completion of this survey, will provide a definitive answer to
this controversial issue.

\subsection{The size - velocity relation}

The bottom panel of Fig.\,3 shows the relation between the disc scale
length and the maximum rotation velocity for both {\small MASSIV}
rotating galaxies and samples from the literature at different
redshifts.

The distribution of the MASSIV rotators in this plot (blue-coded
circles) is consistent with there being no dependence between the disc
size and the stellar mass. For the reasons explained in Sect.\,5.1, we
discuss hereafter the size-velocity relation using the slope
determined for local galaxies. We observe a shift of
$-0.12\pm0.05$~dex in the relation at $z\sim 1.2$ with respect to the
local I-band size-velocity distribution computed by
\citet{dutton2007}. As reported by \citet{puech2007}, different
calibrations proposed to describe this correlation show very similar
results, despite the different quantities used to trace the galactic
properties. Puech et al. (2007) found some IMAGES rotating galaxies at
$z\sim 0.6$ with a relatively lower disc scale length at fixed
velocity than local rotators (cf. blue circles in their Fig.\,3),
although the majority of the galaxies lies on the distribution of
local rotating discs. We fitted their rotating galaxies with a
relation showing a marginal evolution of $-0.08\pm0.06$ from $z\sim
0.6$ to $z\sim 0$.

At $z\sim 2$, the results differ between authors. A systematic
agreement with our measurements is observed at higher redshift when
adopting the half-light sizes computed by \citet{nfs09}. The
evolution at $z\sim2$ shows even smaller sizes at a given maximum
rotation velocity than lower redshift galaxies with an offset of
$-0.22\pm 0.06$~dex from $z=2$ to $z=0$ (cf. Dutton et al. 2011), and
$-0.10$~dex to $z\sim 1.2$. At $z\sim 2$, a different trend is
observed by Bouch{\'e} et al. (2007) combining data from the SINS
survey and \citet{courteau1997} at $z=0$. Bouch{\'e} et al. (2007) found 
no evolution in the zero-point of the size-velocity relation from $z=2$ to 
$z=0$. Dutton et al. (2011) attributed this discrepant result to an inconsistency 
in the SINS half-width half-maximum size reported in Cresci et al. (2009) and
interpreted as exponential disc scale-lengths.

The size-velocity relation shows at all cosmic times a large scatter
(of the order of $\sim 0.2$~dex). The scatter reported in previous
analyses at high-$z$ are likely due to the difficulties in discerning
kinematically complex systems, which translate into uncertainties in
the estimation of the characteristic radius of the system. While we
are not unaffected by these uncertainties, the statistical trend
observed between the size of our ionized emission maps and our radius
of the stellar continuum in the broad-band images of rotating
galaxies, ensures that we are able to derive reliable quantities. The
local relations studied by e.g., \citet{courteau1997} and
\citet{mathewson1992} and more tightly constrained in this respect,
also show a large scatter in this relation suggesting a physical
meaning of this dispersion.

\begin{figure}[ht!]
\begin{center}
\includegraphics[width=6.9cm,angle=0]{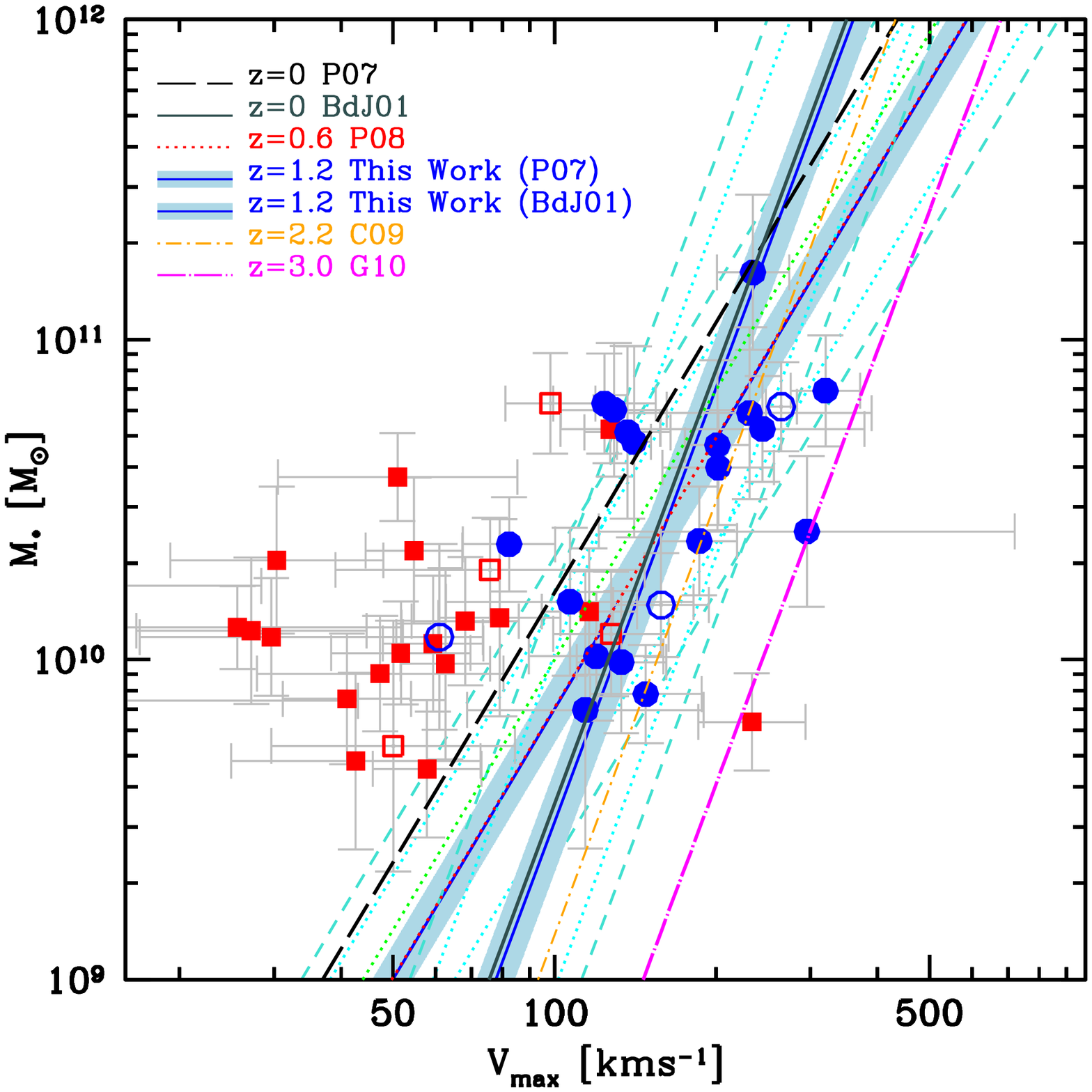} 
\includegraphics[width=6.9cm,angle=0]{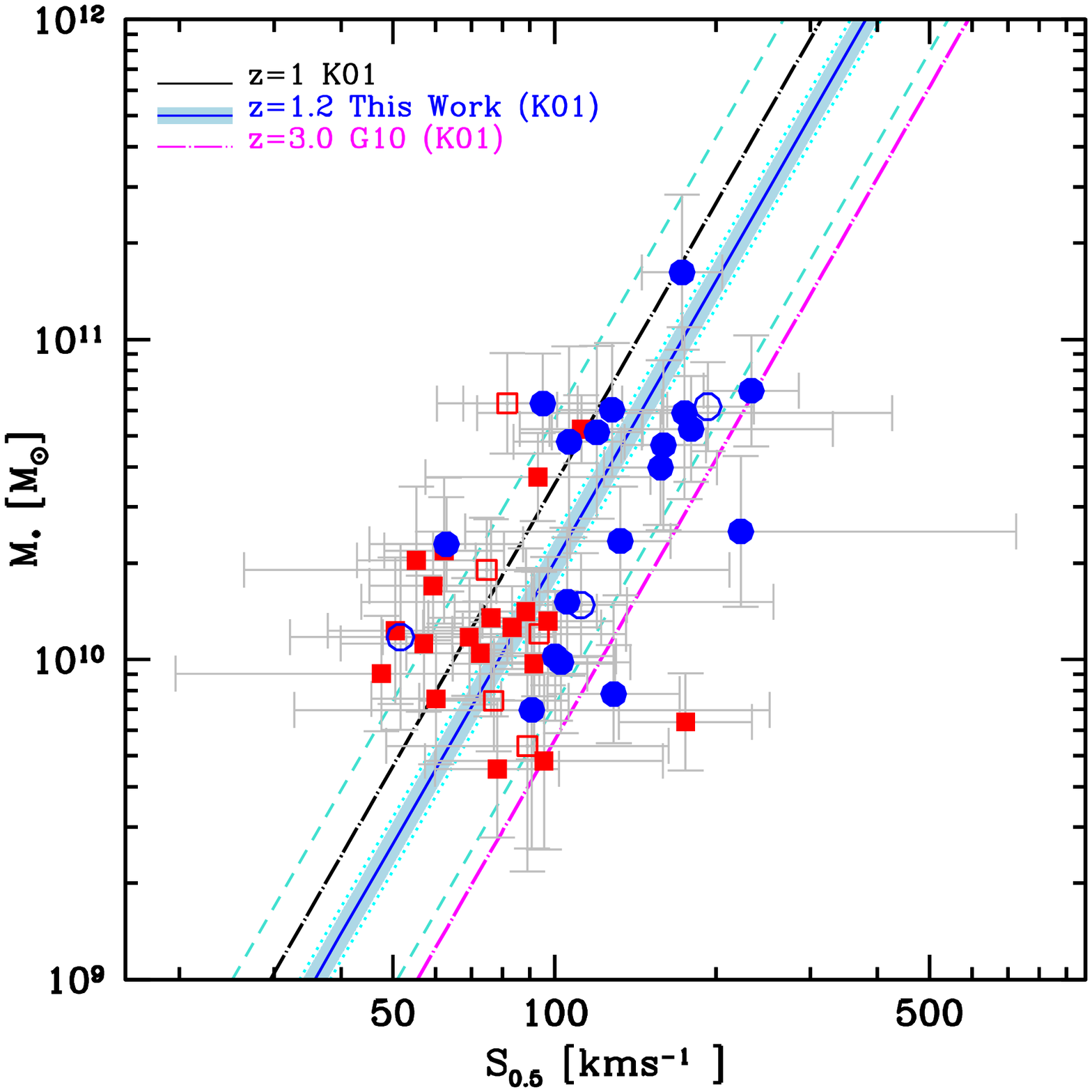} 
\end{center}
\caption{({\bf Top}) The stellar mass TF (smTF) relation at $z\sim
  1.2$ based on the {\small MASSIV} rotating galaxies. Symbols are as in
  Fig.\,1. Errors in the velocities are computed using GHASP
  simulations to account for various uncertainties. The smTF for the
  {\small MASSIV} rotators is calibrated using the local slope defined
  by either Pizagno et al. (2007, P07) or Bell \& de Jong (2001, BdJ01)
  (dashed black line and solid black line, respectively). The
  solid-blue lines are the best-fit to {\small MASSIV} rotating
  galaxies and the cyan shaded area shows the $1\sigma$ error in the
  zero-point parameter (the intrinsic and total scatter is plotted as
  cyan dotted line and cyan dashed line, respectively). Other fits
  are those of \citet{puech08} (P08, $z\sim 0.6$, dotted red line),
  \citet{cresci09} (C09, $z\sim 2.2$, short-dashed-dotted orange
  line), and \citet{gnerucci2011} (G11, $z\sim 3$, long-dashed-dotted
  magenta line). ({\bf Bottom}) The relation between the stellar mass
  and the contribution of both ordered and chaotic motions ($S_{05}
  =\sqrt(0.5 \times v^2_{rot}+\sigma^2_0)$, Kassin et al. 2001, K01)
  is shown. The cyan shaded area shows the $1\sigma$ error on the
  zero-point parameter, the cyan dotted line and cyan dashed line are
  the intrinsic and total scatter, respectively. We overplot the
  best-fit relation of \citet{kassin2007} at $z\sim1$ (K01, dot-dashed
  black line) and \citet{gnerucci2011} at $z\sim3$ (G11, dot-dashed
  magenta line). Symbols are as in Fig.\,1. \label{fig:smtf}}
\end{figure}

\subsection{The mass - velocity relation}

In the top panel of Fig.\,4, we show the correlation between the
stellar mass and the rotation velocity at $z\sim 1.2$ of the {\small
  MASSIV} galaxies. Rotating galaxies are plotted with blue-coded
circles. They are the sole class of galaxies included in our best-fit
procedure. We plot non-rotating galaxies with red-coded, square
symbols. The errors in the velocities are computed using simulations
in a local sample to account for various uncertainties (see for
details \citealt{epinat2008}). The errors on the stellar mass are
described in \citet{contini11}. As we did in the previous section,
given the relatively limited size of our data set at this stage of the
survey we fixed the slope to that of the local calibration. To make a
fair comparison with the published data at different redshifts, and to
allow for future comparisons, we fitted our data with the two most
widely used calibrations. In particular, we calibrated the stellar
mass TF relation (smTFR) with the local slope by \citet{pizagno07} to
compare with the IMAGES sample at $z\sim 0.6$ by
\citet{puech08,puech10}. The local calibration by \citet{pizagno07}
based on a representative galaxy subsample extracted from the SDSS and
revised by \citet{hammer07} is plotted in Fig.\,\ref{fig:smtf} with
dashed, black line. The best-fit relation of \citet{puech08} is
plotted with a dotted red line. We calibrated the relation with the
local slope defined by \citet{bdj01} to compare with the SINS sample
at $z\sim 2.2$ by \citet{cresci09} and with the LSD/AMAZE galaxies at
$z\sim 3$ by \citet{gnerucci2011}. The local calibration of
\citet{bdj01} is plotted in Fig.\,\ref{fig:smtf} with a solid, black
line. The best fit by \citet{cresci09} is shown with a
short-dashed-dotted orange line and that by \citet{gnerucci2011} with
a long-dashed-dotted magenta line. The Bell \& de Jong (2001)
calibration is based on the local sample of \citet{ver01}, who
selected predominantly low-mass, gas-rich galaxies observed with the
21cm {\small HI} emission \citep{puech10, hammer07}. The resulting
relation is steeper than the Pizagno et al. (2007) reference. The two
relations are almost equivalent around a velocity of $\sim$
200~\kms\ and a stellar mass of $\sim 10^{11}$~\msun, thus samples
with these galactic properties are not substantially affected by the
choice of the calibration. The vast majority of our {\small MASSIV}
galaxies are located below these values. In consequence, adopting one,
or the other calibration affects the strength of the evolution in the
velocity-mass relation. For each calibration, we computed the
zero-point of the correlation, its error, its intrinsic, and total
scatter. All the above-mentioned studies have been opportunely
rescaled to account for our Salpeter IMF.

For rotators, we obtain an evolution from $z=0$ in stellar mass of the
smTFR zero point of $-0.36\pm0.11$~dex, using as reference the SDSS
subsample by \citet{pizagno07} (with an intrinsic scatter of
$\sigma_{intr}=0.32$~dex and a total scatter of
$\sigma_{tot}=0.48$~dex). Using the \citet{bdj01} calibration, we
instead obtain a zero-point at $z\sim 1.2$ that is consistent with no
evolution since $z\sim 0$ (or $-0.05\pm0.16$~dex, $\sigma_{intr}=0.52$
and $\sigma_{tot}=0.72$). The solid, blue lines are the best-fit
relation to {\small MASSIV} rotators and the cyan shaded area shows
the $1\sigma$ error on the zero-point parameter (the intrinsic and
total scatter is plotted as cyan dotted and cyan dashed line,
respectively).

The shift in the zero point of the smTFR with respect to $z=0$
computed by \citet{puech08,puech10} at $z\sim0.6$ is surprisingly high
($-0.34$~dex in stellar mass) relative to our results at $z\sim
1.2$. The evolution in the smTFR computed using IMAGES dataset is very
similar (within the errors) to the evolution we report at higher
redshift ($-0.36\pm0.11$~dex at $z \sim 1.2$). There are important
issues to be taken into account when interpreting these values. The
original evolution claimed by \citet{flores06} for the same data-set
was consistent with no or marginal evolution. Other authors, using
traditional slit spectroscopy, reached the same qualitative conclusion
of no/mild evolution up to redshift $z\simeq1$ \citep[e.g.,][]{
  weiner2006, conselice2005, fernandezlorenzo2010}. The IMAGES
galaxies have large seeing and coarse spatial resolution owing to the
characteristics of the FLAMES instrument \citep{puech10}. Furthermore,
the {\it lack} of evolution in the {\it baryonic} (star+gas)
Tully-Fisher relation reported by \citet{puech10} using the same
IMAGES sample and a gas fraction at $z\sim 0.6$ that is similar to the
values of nearby galaxies, implying that there has been a minimal, if
not negligible evolution of the stellar-mass TFR
\citep{dutton2011}. All these points therefore suggest an
overestimation of the evolution of the stellar Tully-Fisher relation
by \citet{puech08} and an evolution of the smTF relation of IMAGES
data of the order of $-0.1$~dex only, instead of the estimated
$-0.34$~dex \citep{dutton2011}.

If we compare our best fit with the Bell \& de Jong (2001) calibration
at $z\sim 1.2$ for the four SINS galaxies by Cresci et al. (2009) at
similar redshifts, it is consistent with no evolution of the zero
point. These four galaxies are very massive ($\sim 10^{11}$\msun) with
velocities around $\sim 250$\,\kms\ where the two calibrations give
similar estimates. At higher redshift, Cresci et al. (2009) found a
shift in the stellar masses at fixed velocity of $-0.41\pm0.11$~dex
from the $z=0$ Bell \& de Jong (2001) calibration. Similar evolution
is observed with the calibration of Pizagno et al. (2007), or
$-0.44\pm0.11$~dex (Cresci priv. comm.). The absence of any
discrepancy when taking two different calibrations is expected giving
the properties of the analysed SINS galaxies. They are very massive
and fast rotators and are located at the intersection between the two
local relations in the stellar mass - velocity diagram. Gnerucci et
al. (2011) analysed the data set of LSD/AMAZE galaxies to discover
evidence of the possible build-up of the stellar mass TF relation at
$z\sim 3$. Despite the large scatter in the correlation ($\sim
1.5$~dex), they report a statistical shift in the zero point relative
to their local relation, of $-1.29$~dex (a very fast evolution,
$-0.88$~dex, between $z\sim 2.2$ and 3 compared to more recent cosmic
epochs).

\medskip Galaxies at high redshifts have a larger velocity dispersion
and a higher fraction of turbulent motions \citep[e.g.,][]{genzel08}.
To account for the large velocity dispersion that contributes to
galactic kinematics at increasing redshift, we plot in the bottom
panel of Fig.\,\ref{fig:smtf} the relation between the stellar mass
and the $S_{05}$ index. The $S_{05}$ index is defined as
$S_{05}=\sqrt(0.5 \times v^2_{rot}+\sigma^2_0)$ to account for the
contributions of both ordered and turbulent motions \citep{kassin2007,
  covington10}.

The cyan shaded area shows the $1\sigma$ error in the zero-point
parameter, the cyan dotted and cyan dashed lines are the intrinsic and
total scatters, respectively. The diagonal lines represent the
best-fit relations of \citet{kassin2007} at $z\sim1$ (dot-dashed black
line) and of \citet{gnerucci2011} at $z\sim3$ (dot-dashed magenta
line) calibrated using the slope of \citet{kassin2007}. We found a
tighter correlation that has a smaller scatter than the classical smTF
relation. Physically this smaller scatter in the distribution
reinforces the results on the importance of chaotic motions to
the galactic kinematics, and the larger contamination encountered in
traditional long-slit spectroscopy.

Using the $S_{05}$ index, the rotators and non-rotators are located
now in the same locus (bottom panel of Fig.\,4). The $z\sim 1.2$
smTF$_{S05}$ is log(M$_{\star}$)= 4.46$\pm0.07$ + 2.92 $\times$
log(S$_{05}$) (with $\sigma_{intr}$=0.08 and $\sigma_{tot}$=0.45) when
galaxies of both classes are included. If only rotating galaxies are
considered (this is the correlation plotted in the figure), the
relation can be written as log(M$_{\star}$)= 4.26$\pm0.10$ + 2.92
$\times$ log(S$_{05}$) (with $\sigma_{intr}$=0.13 and
$\sigma_{tot}$=0.43). We therefore report a shift in stellar mass of
$-0.25$~dex for the rotating {\small MASSIV} sample (or $-0.45$~dex
including both dynamical classes) relative to the \citet{kassin2007}
galaxies at $z\sim 1$ and of $+0.35$~dex (or $+0.55$~dex) relative to
the \citet{gnerucci2011} galaxies at $z\sim3$.

\medskip Figure\,5 shows the relation between the baryonic mass and
the velocities (top panel) and the ordered+chaotic motion components,
quantified using the $S_{0.5}$ index (bottom panel). Based on our
results, we infer that there has been no evolution which should be
compared with the conclusions reached by \citet{mcgaugh05} in the
local Universe implying a very marginal evolution of the content of
the gaseous component with respect to nearby galaxies. Previous
results on the baryonic TF relation using a data-set of integral field
survey were derived by Puech at al. (2010), who found no significant
evolution compared to the local Universe. 

These authors claimed that gas was already in place at $z\sim 0.6$
with no need to advocate external gas accretion since the gas was
already bound to the gravity well of galaxies. Other interpretations
are possible: if the predicted increases with cosmic time in both
galactic velocities and masses occur according to the scaling
relations, as predicted by LCDM models, any evolution should be
observationally recognized.

\begin{figure}[ht!]
\begin{center}
\includegraphics[width=7.9cm,angle=0]{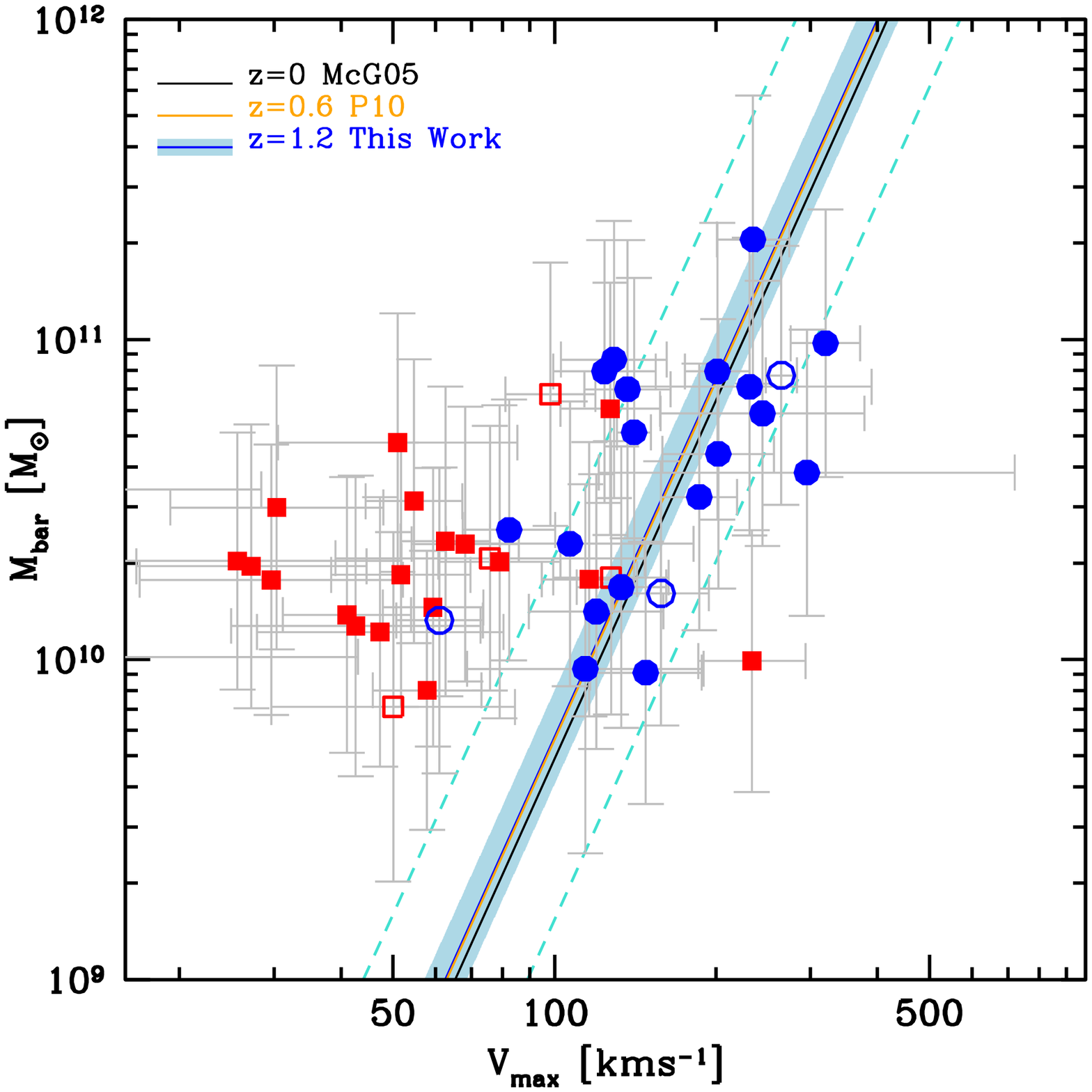}
\includegraphics[width=7.9cm,angle=0]{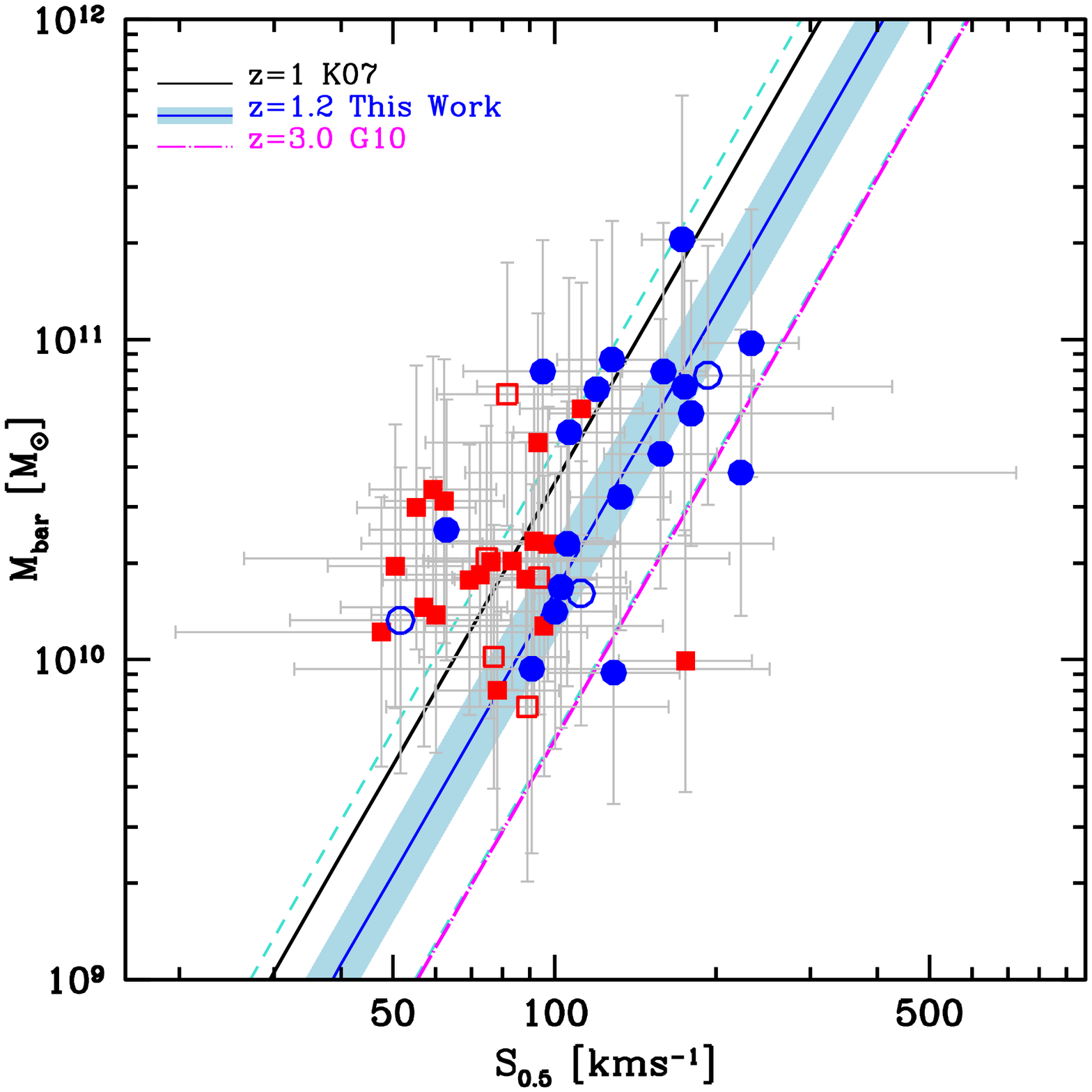}
\end{center}
\caption{ The baryonic TF (bTF) relation versus the maximum velocity
  (top panel) and versus the $v+\sigma$ (or $S_{0.5}$) index (bottom
  panel). We overplot the best-fit relation of rotating galaxies (blue
  symbols) using the local slopes by \citet{mcgaugh05} (plotted as
  solid black line in the top panel, McG05) and by \citet{kassin2007}
  at $z \sim 1$ (plotted as solid black line in the bottom panel,
  K07). The other fit in the top panel is by \citet{puech10} (P10,
  $z\sim 0.6$, solid, orange line) and in the bottom panel by
  \citet{gnerucci2011} ($z\sim 3$, long-dashed-dotted magenta line,
  G11). The cyan shaded area shows the $1\sigma$ error on the
  zero-point parameter (the intrinsic and total scatter is plotted as
  cyan dotted line and cyan dashed line, respectively). Symbols are 
  as in Fig.\,1.}
\end{figure} 

\begin{figure}[ht!]
\begin{center}
\includegraphics[width=8cm,angle=0]{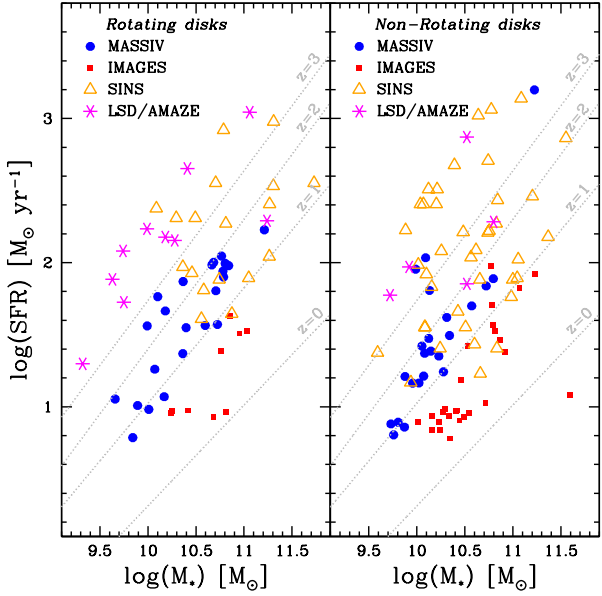}
\end{center}
\caption{Star formation rate versus stellar mass for rotators (left
  panel) and non-rotators (right panel) in the {\small MASSIV}
  ($1<z<1.6$, blue circle), IMAGES ($z\sim0.6$, red square), SINS
  ($z\sim2.2$, orange triangle), and LSD/AMAZE samples ($z\sim3$,
  magenta asterisk). The grey dotted lines show the empirical main
  sequence of star-forming galaxies following \citet{bouche10} at
  different redshifts. All quantities have been rescaled to account
  for our Salpeter IMF.} \end{figure}

 \section{Discussion and conclusions}
 \label{sec:conclusion}
 
We have explored the relationships between galaxy size, mass, and
internal velocity of the statistically representative {\small MASSIV}
sample of 45 galaxies with near-infrared resolved kinematics at $1 <
z< 1.6$, an unbiased sample of high-$z$ star-forming galaxies
\citep{contini11}.

Before the advent of the three-dimensional integral field
spectroscopy, the possible sources of discordance when using these
quantities included observational difficulties and selection effects
\citep{portinari07}. The larger fraction of morphological and
kinematic disturbances in high-$z$ galaxies compared to nearby objects
may bias the conclusions about the real evolutionary effects. However,
systematics and subtle selection effects also need to be taken into
account when comparing with recent near-infrared resolved kinematics
surveys. For example, the observations of the OSIRIS sample
\citep{law07, law09, wright2008} were assisted by adaptive optics to
reach very high angular resolutions, but probing only the highest
surface brightness emission. With this observing mode, only smaller
characteristic H$\alpha$ radii ($\sim 1.3$~kpc) than those typically
probed ($\sim 3$~kpc) in other similar surveys at comparable redshifts
could be studied. The IMAGES survey \citep{hammer07, puech08, puech10}
samples both star-forming and more quiescent galaxies, but the coarse
spatial resolution of the FLAMES instrument requires meticulous
correction methods to extract the maximum velocity. The SINS sample is
the largest collection of galaxies of its kind at high-$z$, but is
extracted from surveys with different selection criteria. The SINS
sample used to analyse the relationship between stellar mass and
velocity, represents galaxies detected with the highest S/N: they are
the most luminous/massive, most rapidly rotating, and most highly
star-forming galaxies.

In the left panel of Fig.\,6, we present the implications of these
observational findings for the physical interpretation, where the star
formation rate of rotating discs obtained by SED fitting in resolved
kinematics surveys is plotted as a function of the stellar mass. The
class of non-rotating/merging systems is shown in the right panel of
the same figure. The main sequence of star-forming galaxies defined by
\citet{bouche10} is drawn at different redshifts. The IMAGES rotators
lie below the sequence defined by the {\small MASSIV} rotators that
was consistently rescaled to star formation rate and stellar mass at
their later cosmic epoch (but covering a large interval of galactic
properties), whereas similarly high-$z$ LSD/AMAZE rotating discs stay
above the sequence. The SINS rotators are highly star-forming objects
and very massive, thus coincide with the massive, star-forming tail of
the {\small MASSIV} galaxy distribution. The SINS properties can
explain the mostly absent scatter in the Tully-Fisher relation by
Cresci et al. (2009) relative to our scatter in the same relation
(cf. our Fig.\,4 with their Fig.\,5). In contrast to our expectations,
the difference in SFR between {\small MASSIV} and SINS galaxies cannot
be attributed to the higher redshift and/or to the limits imposed on
the minimum line flux to ensure detection \citep{contini11}. In this
figure, there is a larger dispersion in SINS rotators around the main
sequence compared to other samples. This may be the consequence of at
least two combined effects. The first cause may be the heterogeneous
SINS selection. The second cause may be physical, and related to an
unexpectedly larger incidence of internal/external mechanisms
(inflows/mergers, outflows/feedback) acting on both the
spectrophotometric and dynamical properties of rotating discs. While
we cannot discard the second hypothesis, it is unlikely that the same
mechanisms acting on SINS galaxies at $z \sim 2$ are not at play a few
Gyr later in {\small MASSIV} galaxies (assuming both samples are
representative of the entire rotating disc population). Nevertheless,
we can speculate that we witness the same phenomena acting with
different efficiencies in the various physical regimes (of galactic
stellar mass, star-formation rate, etc.). Furthermore, comparing the
SFR of rotators (left panel in Fig.\,6) and non-rotators (right panel)
for the SINS and {\small MASSIV} surveys, it appears that non-rotating
systems display similar ranges of SFR at $z\sim1$ and $z\sim2$, while
this is not the case for rotating systems that exhibit higher SFR at
$z\sim2$ than at $z\sim1$. Our results highlight the importance of the
disc formation to the SFR mechanics between $z\sim2$ and $z\sim1$. In
summary, the galaxy datasets that have become available so far (even
those taking advantage of resolved spectroscopy) are spectroscopic
surveys that pass through a natural pre-selection from one, or several
photometric catalogues. As a consequence, the comparison between
samples remains difficult. At present, firm conclusions based on
comparisons with other surveys that are affected in different ways by
systematics in the galaxy properties determinations (e.g., different
selection effects, quantities obtained at different wavelengths, and
different assumptions), should be formulated with caution.\medskip

In this paper we have attempted a statistical analysis of our data
set. We have analysed a large representative catalogue of galaxies
with well-known selection functions, that is based on flux-limited
multiple fields of a homogeneous redshift survey observed at higher
spectral resolution, but with sufficient spatial resolution probing
typical scales of $\sim 5$~kpc. Our results are the following.
 
The {\small MASSIV} galaxies show a correlation between dynamical and
stellar mass with an offset between the two that can be interpreted as
the gas mass being of the order of 20-25\% of the stellar mass at
$z\sim 1.2$. A similar fraction of gas mass is obtained using the
Kennicutt-Schmidt formulation. We have found that this fraction of gas
mass has evolved mildly for the past 8~Gyr, a result that is
reinforced by the lack of a statistically significant evolution of our
baryonic Tully-Fisher relation. In contrast, our finding of a
disagreement in the gas estimates using two independent approaches for
non-rotating galaxies, suggests different properties of dark matter
halos between rotators and systems not supported by rotation at
high-$z$, as is prescribed for nearby objects. The stellar mass in
{\small MASSIV} galaxies is correlated with the incidence of being
either a rotator or a non-rotator, with non-rotating galaxies
containing a smaller amount of stellar mass at fixed gas content and
dynamical mass, but not at fixed star-formation rate. Furthermore,
non-rotators have more compact stellar components than rotators, but
not for their gas components.

We have derived correlations between galaxy size, mass, and internal
velocity, as observed in the local Universe and indicated by previous
results, but the size - mass and size - velocity relations are weak at
$1 < z < 1.6$. While individual galactic properties are found to
evolve with time as advocated by the most recent results of large
galaxy surveys, the marginal change observed in the relationships
linking the three fundamental quantities analysed in this work $-$
size, mass, and internal velocity $-$ suggests that there is some
evolution in these relations. This picture agrees well with the
cosmological N-body/hydrodynamical simulations by e.g.,
\citet{portinari07} and \citet{firmani2009}. We do not confirm the
hypothesis of a strong, positive evolution in the size - stellar mass
and size - velocity relations, where discs were found to be evenly
smaller with look-back time at fixed stellar mass or velocity. Our
results do not imply that there has been any unusual evolution in the
galactic spin similar to that reported by \citet{bouche07}. However,
we have discovered a large spread in the distributions, which is
significantly smaller for non-rotating galaxies when including the
contribution of turbulent motions (with the $S_{05}$ index) in the
relations involving the velocities. This result disagrees with the
hypothesis that large scatters in the low-velocity regime are due to
disturbed or compact objects, as suggested by
\citet{covington10}. Restricting our analysis to regularly rotating
discs that sample a large range of stellar masses and sizes, we have
found that there is a persistent scatter. Its origin is likely
intrinsic, and possibly caused by complex physical mechanism(s) at
work in our stellar mass/luminosity regime and redshift range. At the
present stage of the {\small MASSIV} survey and given the typical
scale ($\sim 5$~kpc) used to trace the resolved kinematics, we have
been unable to investigate further the mechanism(s) driving these
galactic properties. It is still a matter of debate as to whether we
trace more turbulent discs where star formation is the energetic
driver \citep[e.g.,][]{lehnert2009,green10}, or we are influenced by
the infall of accreting matter, and/or large collisional clumps
\citep[e.g.,][]{law07, law09,genzel2011}, and/or merger events
\citep[e.g.,][]{hammer2011}.

More sophisticated analyses will be possible on the completion of our
survey, doubling the statistics with galaxies in our highest ($z>1.2$)
redshift range. However, some problems will only be able to be
addressed with larger-scale surveys of kinematically resolved
galaxies. Future surveys should be able to assess firm conclusions
about the cosmic evolution of the slope of the size/mass/velocity
fundamental relations of discs at $z>1$. There is also the pressing
need to establish the imprint of the environmental processes at work
in various stellar mass regimes. The stellar component and ionized gas
content of galaxies of course offer only a partial view of galaxy
formation and evolution, which must be supplemented with the
properties of the cold/molecular phase of the interstellar
component. In the future, we hope to comprehend under which
conditions, and how efficiently, the gas is converted into stars
through cosmic epochs. The relation between the galactic properties
probed in our survey and the characteristics of the gas reservoir will
offer a crucial step to link the data to models of galaxy formation.

\section{Acknowledgments}

DV acknowledges the support from the INAF contract
PRIN-2008/1.06.11.02. This work has been partially supported by the
CNRS-INSU and its Programme National Cosmologie-Galaxies (France) and
the French ANR grant ANR-07-JCJC-0009.

\bibliographystyle{aa} \bibliography{18453_biblio}

\end{document}